\documentclass[11pt,a4paper]{article}

\usepackage{graphicx}
\usepackage{epsfig}
 \usepackage[titletoc,toc,title]{appendix}
\usepackage{hyperref}

\usepackage[dvips]{color}

\usepackage{graphicx,epsf}
\usepackage{bm}
\usepackage{amsmath}
\usepackage{pifont}
\usepackage{amssymb}

\newcommand{\Nr}{N_r}
\newcommand{\Ntheta}{N_\theta}
\newcommand{\Nphi}{N_\varphi}
\newcommand{\Nvpar}{N_{v_\parallel}}
\newcommand{\Nmu}{N_\mu}
\newcommand{\rmin}{r_{\rm min}}
\newcommand{\vpar}{v_\parallel}

\newcommand{\dd}{\:{\rm d}}
\newcommand{\jacobv}{{\cal J}_{\rm v}}

\newcommand{\Omegas}{\Omega_s}
\newcommand{\pot}{\phi}
\newcommand{\vecv}{\mathbf{v}}
\newcommand{\vecx}{\mathbf{R}}
\newcommand{\vGpar}{\vpar}

\makeatletter
\def \yb #1{ %
}
\def \yb #1{ %
\boldsymbol{#1} %
}

\newcommand{\vecb}[1]{{\bf #1}}
\def \d {\mathrm{d}}

\begin{document}

\begin{center}
{\large{\textbf{Cross-code gyrokinetic verification and benchmark on the linear collisionless dynamics of the geodesic acoustic mode.}}}\\
\vspace{0.2 cm}
{\normalsize {A. Biancalani, A. Bottino, C. Ehrlacher, V. Grandgirard, G. Merlo, I. Novikau, Z. Qiu,\\ E. Sonnendr\"ucker, X. Garbet, T. G\"orler, S. Leerink, F. Palermo, D. Zarzoso.}}\\
\vspace{0.2 cm}
\small{\url{http://www2.ipp.mpg.de/~biancala/}\\
\vspace{0.7cm}}
\end{center}

\begin{abstract}
The linear properties of the geodesic acoustic modes (GAM) in tokamaks are investigated by means of the comparison of analytical theory and gyrokinetic numerical simulations. The dependence on the value of the safety factor, finite-orbit-width of the ions in relation to the radial mode width, magnetic-flux-surface shaping, and electron/ion mass ratio are considered. Nonuniformities in the plasma profiles (such as density, temperature, or safety factor), electro-magnetic effects, collisions and presence of minority species are neglected. Also, only linear simulations are considered, focusing on the local dynamics.
We use three different gyrokinetic codes: the lagrangian (particle-in-cell) code ORB5, the eulerian code GENE and semi-lagrangian code GYSELA.
One of the main aims of this paper is to provide a detailed comparison of the numerical results and analytical theory, in the regimes where this is possible. This helps understanding better the behavior of the linear GAM dynamics in these different regimes, the behavior of the codes, which is crucial in the view of a future work where more physics is present, and the regimes of validity of each specific analytical dispersion relation.
\end{abstract}

\section{Introduction}
\label{sec:intro}

Turbulence in tokamak plasmas is often observed accompanied by zonal, i.e. axisymmetric, radial electric fields, giving rise to zonal poloidal flows. Two kinds of zonal flows are identified: zero-frequency zonal flows (ZFZF)~\cite{Hasegawa79,Rosenbluth98,Diamond05} and oscillating zonal flows, named geodesic acoustic modes (GAM)~\cite{Winsor68,Zonca08}
GAMs have mainly zonal polarization of the perturbed electric field, i.e. n=0, m=0, with n and m being respectively the toroidal and poloidal magnetic field, and n=0, m=1 perturbed density. Their characteristic frequency is of the order of the sound frequency $\omega \sim c_s /R_0$ (where $c_s=\sqrt{T_e/m_i}$ is the sound speed and $R_0$ is the tokamak major radius, with $m_i$ being the ion mass and $T_e$ the electron temperature).
The importance of understanding the dynamics of these zonal structures in tokamaks is due to their nonlinear interaction with turbulence, being crucial for its saturation~\cite{Scott92,Miyato04,Scott05,Angelino06,Conway11}. As an example of some recent experimental observations of GAMs in ASDEX Upgrade, see Ref.~\cite{Simon16}.

In this work, we focus on the linear properties of GAMs, and investigate the dependence on the safety factor, the effect of the finite-larmor-radius (FLR) and finite-orbit-width (FOW) of the ions in relation to the GAM radial width, the effect of the magnetic-flux-surface shaping, and of the electron/ion mass ratio. Nonuniformities in the plasma profiles (such as density, temperature, or safety factor), magnetic effects, collisions and effects of plasma minorities (such as bulk ion impurities or energetic ions) are neglected here. Our investigation is done by means of analytical theory and numerical simulations.

A great progress in the analytical investigation of the linear GAM dynamics has been achieved, starting with its first estimate in ideal MHD~\cite{Winsor68}, then in kinetic theory by neglecting the effect of the FOWs of the passing ions~\cite{Zonca96,Zonca08}, then including it to the first order~\cite{Zonca98,Sugama06,Sugama08,Zonca08}, then including it to higher orders~\cite{Xu08prl,Qiu09ppcf}, and then including the effect of the flux surface shape~\cite{Gao09,Gao10pop}.
All such analytical models neglect the effect of the finite mass of the electrons. In fact, all analytical theories derived so far treat the $m\ne 0$ component of the electrons as adiabatic (whereas the m=0 component of the electron density perturbation is imposed to zero). We will refer to this model for treating the electrons as ``adiabatic''. The importance of having an analytical description is twofold.
On the one hand, it allows a direct understanding of the physical mechanisms leading to the each different effect under investigation. On the other hand, it allows a detailed linear verification process of the numerical tools, which is at the basis of the development of gyrokinetic codes aimed at a rigorous turbulence investigation.

Many numerical investigations of the linear GAM dynamics and comparison with analytical theory or benchmark among codes have been carried out in the last few decades, most of which treating the electrons as adiabatic.
As a non-extensive list of example, we mention here simulations performed with the gyrokinetic codes GTC~\cite{Rewoldt98,Lin00}, ORB5~\cite{Angelino08} (where the effect of the elongation was studied), TEMPEST~\cite{Xu08prl} (where the effect of high-order terms of the finite ion orbit width was studied), GYRO~\cite{Hager09} (where the effect of finite orbit width and the application to the radial velocity in the large-q limit was studied), GYSELA~\cite{Zarzoso12}, ELMFIRE~\cite{Heikkinen08}, and GENE with GKW~\cite{Merlo16}.
In particular, a first verification of ORB5 against analytical theories, for circular geometries and low values of $k_r \rho_i$, was started in Ref.~\cite{Biancalani14}.
A numerical study of the effect of kinetic electrons in circular plasmas  has been described in Ref.~\cite{Zhang10}.

In this paper, we aim at performing a comprehensive cross-code verification and benchmark of several gyrokinetic codes, in different regimes.
We perform numerical simulations with three different gyrokinetic codes, adopting equivalent physical models for the dynamics of the ions (which is the most basic species to be investigated for the physics of sound waves, and therefore of GAMs), but solving the model equations in three different ways:
the Lagrangian (i.e. particle-in-cell) code ORB5~\cite{Jolliet07,Bottino11,Bottino15JPP}, the Eulerian code GENE~\cite{Jenko00,Goerler11}, and the semi-Lagrangian code GYSELA~\cite{Grandgirard06,Grandgirard16}. All these codes solve the ion dynamics based on the gyrokinetic equations (see for example Ref.~\cite{Antonsen80,Frieman82,Littlejohn83,Hahm88} for some early derivations, or Ref.~\cite{Brizard07} for a recent comprehensive review).
Some of the differences in the physical models of these codes are the treatment of the electrons and the possibility of investigating the dynamics in elongated magnetic equilibria.

The gyrokinetic model has been also adopted in the past for deriving analytical dispersion relations for linear collisionless GAMs in various regimes (small or large values of safety factors, negligible, small or moderate values of normalized radial wave-number, small or moderate elongation of the equilibrium magnetic flux surfaces).
Testing a numerical code, by using it in a particular limit where the analytical solution is known, takes the name of code ``verification''. Testing different numerical codes, which solve the basic equations in different ways, by using them in particular regimes where the hypothesis of the physical models are the same, takes the name of code ``benchmark''.
The verification and benchmark of gyrokinetic codes has been the main goal of an European effort developed in the last three years as an Eurofusion project.
This project focuses on different physical phenomena observed in tokamaks, like for example GAMs, and ion-temperature-gradient instabilities, both crucial actors to be understood in the view of the prediction and control of the turbulent transport in existing tokamak devices and future fusion reactors. This paper in particular, summarises the results of the verification and benchmark on GAMs (see Ref.~\cite{Goerler16} for a description of the benchmark on ITGs).

The paper is organized as follows. The numerical models of the three codes adopted for this study are described in Sec.~\ref{sec:models}. The dependence of the GAM dynamics on the safety factor, the effect of ion FOW and flux-surface elongation is discussed in Sec.~\ref{sec:ad_ele}, where the electrons are treated as adiabatic. The effect of the finite mass of electrons is described in Sec.~\ref{sec:kin_ele}, where the results of numerical simulations with kinetic electrons are shown. The conclusions are summarised in Sec.~\ref{sec:conclusions}. The investigation of the numerical convergence of the codes is presented in the appendices.

\section{The models}
\label{sec:models}

The choice of the model for the investigation of the dynamics of GAMs is dictated by their specific spatial and temporal characteristic scales. In particular, GAMs are zonal oscillations (i.e. with toroidal and poloidal mode numbers equal to zero)  with radial wavelength bigger than (or in same case of the order of) the ion Larmor radius, and time scales of the order of the sound time $\sim 2\pi R/c_s$. The basic physics is the one of the sound waves, therefore the MHD description is sufficient for estimating the order of magnitude of the frequency of the mode~\cite{Winsor68}.
Nevertheless, such spatio-temporal scales make the need of a kinetic treatment clear. This is due to the importance of resonances with passing ions, which can determine frequency and damping rate of the modes. Considering the requirements for the spatial scales, and the frequencies being lower than the ion cyclotron frequency, we can easily see that the gyrokinetic model~\cite{Antonsen80,Frieman82,Littlejohn83,Hahm88,Brizard89,Sugama00,Brizard07} is the most appropriate tool.

\subsection{The numerical model of ORB5}
\label{sec:ORB5-model}

ORB5 is a nonlinear gyrokinetic code based on a particle-in-cell (PIC) algorithm.
The basic discretization scheme of a PIC code (also known as ``Lagrangian'' code), for the Vlasov-Maxwell problem, is presented in Ref.~\cite{Birsdall89}.
A PIC code discretizes the distribution function with macro-particles, also known as markers, associated with weights. In a gyrokinetic PIC code, the markers are pushed along the trajectories derived from the gyrokinetic model while the fields are known on a spatial grid and evolved by solving the gyrokinetic field equations either with finite differences or with finite element methods.
The sources (charge density and current density) needed for solving the field equations are calculated by projecting the marker weights on a spatial grid.
In ORB5, the distribution function is decoupled in a background distribution analytically known, while only the perturbation is solved using markers, with a control-variate Monte Carlo method, hystorically known as $\delta$f PIC~\cite{Lee83} (see Ref.~\cite{Bottino15JPP} for a recent overview).

ORB5 was originally developed for electrostatic turbulence studies~\cite{Jolliet07}.  
In the last few years, it has been extended to the electromagnetic, multi-species version within the NEMORB project~\cite{Bottino11,Bottino15JPP}.
In this paper, only the linearized electrostatic model of ORB5 is used. Only collisionless simulations are considered. Although only the local GAM dynamics is of interest in this paper, no flux-tube version of ORB5 exists, therefore only global simulations are considered, and the global effects are neglected (see Ref.~\cite{Palermo17} for an investigation of the global effects with ORB5).
The model equations of ORB5 are derived in a Lagrangian formulation~\cite{Bottino15JPP}, based on the gyrokinetic Vlasov-Maxwell equations of Sugama, Brizard and Hahm~\cite{Sugama00,Brizard07}. 

The gyrocenter trajectories describe the motion of the markers of the kinetic species in phase-space coordinates written in $p_\|$-formalism, $\vecb{Z}_{p_\|}=(\vecb{R},p_\|,\mu)$, i.e. respectively the gyrocenter position, canonical parallel momentum $p_\| = m_s v_\| + (q_s/c) \tilde{A}_\|$ and magnetic momentum $\mu = m_s v_\perp^2 / (2B)$ (with $m_s$ and $q_s$ being the mass and charge of the species).  $v_\|$ and $v_\perp$ are respectively the parallel and perpendicular component of the particle velocity.
The gyroaverage operator is labeled here by the tilde symbol $\tilde{}$ . The gyroaverage operator reduces to the Bessel function $J_0$ if we transform into Fourier space. In all simulations with ORB5 shown in this paper, the gyroaverage is always calculated by considering non-vanishing Larmor radius for the ions, whereas it is calculated with zero argument for the electrons. In other words, finite-Larmor-radius (FLR) effects are retained for ions and neglected for electrons.
The code ORB5 is based on straight-field-line tokamak coordinates. Dirichlet boundary conditions are imposed in the radial direction, while periodicity is assumed in the two angles.
The nonlinear electromagnetic version of the trajectories is~\cite{Bottino15JPP}:
\begin{eqnarray}
\dot{\vecb  R}&=&\frac{1}{m_s}\left(p_\|-\frac{q_s}{c}\tilde{A}_\parallel\right)\frac{\vecb{B^*}}{B^*_\parallel} + \frac{c}{q_s B^*_\parallel} \vecb{b}\times \left[\mu \nabla B + q_s \nabla   \big(\tilde\phi -  \frac{p_\|}{m_s c} \tilde{A}_\| \big) \right]  \\
\dot{p_\|}&=&-\frac{\vecb{B^*}}{B^*_\parallel}\cdot\left[\mu \nabla B + q_s
  \nabla  \big(\tilde\phi -  \frac{p_\|}{m_s c} \tilde{A}_\| \big) \right] \\
 \dot{\mu} & = & 0
\end{eqnarray}
Here, the time-dependent fields are the scalar potential $\phi$ and the parallel component of the vector potential $A_\|$, and $\vecb{B}^*= \vecb{B} + (c/q_s)  \vecb{\nabla}\times (\vecb{b} \, p_\|)$, where $\vecb{B}$ and $\vecb{b}$ are the equilibrium magnetic field and magnetic unitary vector. The linearization of the Vlasov equation is performed by pushing the markers along unperturbed trajectories:
\begin{eqnarray}
\dot{\vecb  R}&=& \frac{p_\|}{m_s}\, \frac{\vecb{B^*}}{B^*_\parallel} + \frac{c}{q_s B^*_\parallel} \vecb{b}\times \mu   \nabla B  \label{eq:trajectories_a} \\
\dot{p_\|}&=&-\frac{\vecb{B^*}}{B^*_\parallel}\cdot \mu \nabla B \label{eq:trajectories_b}
\end{eqnarray}
In this paper, the trajectories given by Eq.~\ref{eq:trajectories_a}, \ref{eq:trajectories_b} are always calculated for the ions (which are always treated kinetically), whereas the electrons can be either treated kinetically (by considering $J_0=1$ and neglecting the electron polarization) or with an ``adiabatic'' model, where the electron gyrocenter density is calculated directly from the value of the scalar potential as~\cite{Bottino15JPP}:
\begin{equation}
n_e(\vecb  R,t) = n_{e0} + \frac{q_s n_{e0}}{T_{e0}} \big( \phi - \bar\phi \big)   \label{eq:adiabatic-electrons}
\end{equation}
where $\bar\phi$ is the flux-surface averaged potential. The quantities with subscript ``0'' refer to the equilibrium, and therefore are functions of the radial coordinate $\rho$ only.

The equation for solving the scalar potential is the gyrokinetic Poisson equation, also known as polarization equation. This is derived from the gyrokinetic Lagrangian of ORB5, using the variational derivation, and imposing that the ExB drift energy of the particles is larger than the field energy (quasi-neutrality condition)~\cite{Bottino15JPP}. The gyrokinetic Poisson equation is~\cite{Bottino15JPP}:
\begin{equation}
 - \vecb{\nabla} \cdot \frac{n_0 m_i c^2}{B^2} \nabla_\perp \phi=  \sum_s \int \d W  q_s \, \tilde{\delta f_s}  \label{eq:Poisson}
\end{equation}
with $n_0 m_i$ being here the total plasma mass density (approximated as the ion mass density), and the summation over the species is performed when the electrons are treated as kinetic, otherwise the electron contribute is given by $-n_e(\vecb  R,t)$.
Here $\delta f = f - f_0$ is the gyrocenter perturbed distribution function, with $f$ and $f_0$ being the total and equilibrium (i.e. independent of time, assumed here to be a Maxwellian) gyrocenter distribution functions. The integrals are over the phase space volume, with $\d W = (2\pi/m_i^2) B_\|^* \d p_\| \d  \mu$ being the velocity-space infinitesimal. The gyrokinetic Poisson equation is solved with a finite element method, by using B-splines in all the spatial directions.

Eqs~\ref{eq:trajectories_a},~\ref{eq:trajectories_b},~\ref{eq:adiabatic-electrons},~\ref{eq:Poisson} are the constitutive equations of the model of ORB5 used in this paper for studying the collisionless electrostatic linearized dynamics of GAMs. In the electromagnetic version, the Amp\`ere equation is also solved for calculating the time evolution of the parallel component of the vector potential $A_\|$, which is neglected in this paper.

\subsection{The numerical model of GENE}
\label{sec:GENE-model}

The Gyrokinetic Electromagnetic Numerical Experiment (GENE) code, is also a nonlinear gyrokinetic code originally developed for electromagnetic turbulence studies in the flux-tube (i.e. local) limit~\cite{Jenko00}, recently extended to its global representation~\cite{Goerler11}. The model of GENE is also based on the gyrokinetic Vlasov-Maxwell equations of Brizard and Hahm~\cite{Brizard07}. Intra- and inter-species collisions (both pitch angle and energy scattering) are implemented. In this paper, only the linearized electrostatic collisionless version of GENE is used. 

GENE is a Eulerian code. In a Eulerian description, the distribution function is not discretized with markers, but it is discretized on a 5D fixed grid in phase-space. The gyrokinetic equation is then solved on this grid for each species $s$. The coordinate system of GENE in the 5D phase space is written in $v_\|$-formalism, $\vecb{Z}_{v_\|} = (\vecb{R},v_\|,\mu)$, i.e. respectively the gyrocenter position, parallel velocity and magnetic momentum. GENE adopts a field-aligned coordinate system  to represent the fluctuation fields in the configuration space of $\vecb{R}$. This coordinate system becomes singular at the magnetic axis which therefore cannot be simulated.
In the local version of the code, the radial direction is Fourier transformed and periodic boundary conditions are applied. In the global version the radial direction is instead treated in real space and Dirichlet boundary conditions are applied. The binormal direction (i.e. perpendicular to the radial direction and to the equilibrium magnetic field) is always Fourier transformed as axisymmetry corresponds to invariance in this direction, and each linear mode corresponds to a toroidal mode number $n$.

The distribution function $f_s$ of each species is evolved accordingly to the gyrokinetic equation in the form~\cite{Goerler11}:
\begin{equation}
 \frac{\partial f_s}{\partial t} + \frac{d \vecb{R}}{d t}\cdot \nabla f_s + \frac{d v_\|}{d t} \frac{\partial f_s}{\partial v_\|} = 0
 \label{eq:gyrokinetic-GENE}
\end{equation}
where the equations of motion of the gyrocenters  are given by~\cite{Goerler11}:
\begin{eqnarray}
\frac{d \vecb{R}}{d t} & = & v_\| \vecb{b} + \frac{B}{B^*_{G\|}} (\vecb{v}_E + \vecb{v}_{\nabla B} + \vecb{v}_c) \label{eq:trajectories_a-GENE}\\
\frac{d v_\|}{d t} & = & - \frac{d\vecb{R}/dt}{m_s v_\|} \cdot \Big( q_s \vecb{\nabla} \tilde\phi + \frac{q_s}{c} \vecb{b} \frac{\partial  \tilde{A}_\|}{\partial t} + \mu \vecb{\nabla}B \Big)\label{eq:trajectories_b-GENE-em}
\end{eqnarray}
Here $\vecb{B}^*_G= \vecb{B} + ( m_s c/q_s)  \vecb{\nabla}\times (\vecb{b} \, v_\|)$, the generalized ExB drift is $\vecb{v}_E = (c/B^2) \vecb{B}\times \vecb{\nabla} (\tilde\phi-(v_\|/c) \tilde{A}_\|)$, the grad-B drift is $\vecb{v}_{\nabla B} = (\mu c / q_s B^2) \vecb{B}\times \vecb{\nabla}B$, and the curvature drift is $\vecb{v}_c = (v^2_\| / \Omega_s) (\nabla\times\vecb{b})_\perp$.
In the electrostatic version of the code, used in this paper, the ExB drift is $\vecb{v}_E = (c/B^2) \vecb{B}\times \vecb{\nabla}\tilde\phi$, and the second term on the right hand side of Eq.~\ref{eq:trajectories_b-GENE-em} is dropped, so that the equation of the time derivative of the parallel component of the velocity takes the form:
\begin{eqnarray}
\frac{d v_\|}{d t} & = & - \frac{d\vecb{R}/dt}{m_s v_\|} \cdot \Big( q_s \vecb{\nabla}\tilde\phi + \mu \vecb{\nabla}B \Big) \label{eq:trajectories_b-GENE}
\end{eqnarray}
The linearization in GENE is done by plugging in the equation of motion in the Vlasov equation and neglecting all the nonlinear terms. Only linear simulations are considered in this paper.

Equation \eqref{eq:gyrokinetic-GENE} is then solved self-consistently with the gyrokinetic Maxwell equations, which in the cases considered here reduce to the gyrokinetic Poisson equation, Eq.~\ref{eq:Poisson}, which is solved for obtaining the scalar potential (whereas the Amp\`ere equation can also solved in case of electromagnetic simulations).
As in the ORB5 code, different models are available for describing each species dynamics. In this paper ions are always assumed to be fully gyrokinetic whereas electrons, depending on the particular case being simulated, are treated either as a second kinetic species or assumed to respond adiabatically. For typical tokamak parameters the Debye length is much smaller than the characteristic wave-length of microinstabilities. The gyrokinetic Poisson equation can thus be reduced to the quasi-neutrality condition, which, having assumed a quasi-neutral background, reads
\begin{equation}
\sum_sq_s\delta n_s=0
\label{eq:qn}
\end{equation}
where $\delta n_s$ indicates the perturbed gyrocenter density of the $s$-th species, obtained from the gyrokinetic model. When all species are treated kinetically, the equation for quasi-neutrality, Eq.~\eqref{eq:qn} can be rewritten as
\begin{equation}
\sum_s\frac{2\pi q_s}{m_s}\int B^*_{G\|}\tilde{\delta f_s}\d v_{\parallel}\d\mu -\frac{q_s^2n_{0s}}{T_{0s}} \left[\phi-\frac{B_0}{T_{0s}}\int\tilde{\tilde{\phi}}\exp \left(-\frac{\mu B_0}{T_{0s}}\right)\d\mu\right]=0,
\end{equation}
while in case of adiabatic electrons it reduces to
\begin{equation}
-\frac{e^2n_{0e}}{T_{0e}}\left(\phi - \bar{\phi}\right) + \sum_{s\neq e}\frac{2\pi q_s}{m_s} \int B^*_{G\|} \tilde{\delta f_s} \d v_{\parallel}\d\mu
-\frac{q_s^2n_{0s}}{T_{0s}}\left[\phi-\frac{B_0}{T_{0s}}\int\tilde{\tilde{\phi}}\exp\left(-\frac{\mu
    B_0}{T_{0s}}\right)\d\mu\right]=0.
\end{equation}

As mentioned above, the linear physical models of ORB5 and GENE are equivalent (see Ref.~\cite{Tronko17} for a detailed discussion on the comparison of the two models), and no difference in the results is expected for the linear collisionless GAM dynamics, depending on that. Nevertheless, the numerical schemes are different. Moreover, the existence of the two representations of GENE, namely the global and the local (i.e. flux-tube) representations, offers the possibility to solve the model equations in two independent ways. As shown in the following sections, no difference is found in the results, for the chosen tests. This means that, for these particular cases, the local dynamics is dominant and independent on the adopted numerical scheme.


\subsection{The numerical model of GYSELA}
\label{sec:GYSELA-model}
Like ORB5 and GENE, the GYrokinetic SEmiLAgrangian code (GYSELA) is also a nonlinear 5D gyrokinetic code \cite{Grandgirard16}.
No linear version exists, therefore nonlinear simulations are considered in this paper, but with sufficiently small initial perturbation, in order to focus on the linear dynamics.
The GYSELA code is dedicated to electrostatic Ion Temperature Gradient (ITG) turbulence with possibility to address transport of impurities. 
Electrons are presently assumed adiabatic but a kinetic version is under development. 
GYSELA is a global full-$f$ flux-driven code which addresses turbulent and neoclassical transports on an equal footing. 

GYSELA is a global code with a toroidal geometry with a simplified concentric circular magnetic configuration. 
Its coordinate system in the 5D space is written as GENE in $v_\|$-formalism, $\vecb{Z}_{v_\|} = (\vecb{R},v_\|,\mu)$ but where $R=(r,\theta,\varphi)$ with $r$ the radial direction and $\theta$ and $\varphi$ the poloidal and toroidal geometric angles.
Boundary conditions are periodic in $\theta$ and $\varphi$ directions.
Non-axisymmetric fluctuations of the electric potential and of the distribution function - i.e $(m, n)\ne (0,0)$ modes - are forced to zero at both radial boundaries of the simulated domain. 
As far as the axisymmetric component is concerned, the value of the potential is prescribed at the outer boundary, while the radial electric field is set to zero at the inner boundary.
No flux-tube version of GYSELA exists, but since in this paper only local physics is concerned, density, temperature and safety factor profiles will be considered constants to minimize the global effects.
GYSELA is a full-$f$ code, namely the back reaction of turbulent transport is accounted for in the time evolution of the equilibrium. In such a framework, the turbulence regime is evanescent if no free energy is injected in the system. 
A flux-driven version of the code is available since 2009 \cite{Sarazin_NF2010}, where the system can be driven by a prescribed volumetric source, versatile enough to allow for separate injection of heat, parallel momentum and vorticity. However in this paper, the temperature and density profiles are constant and therefore we only use the forcing governed by the two equal thermal baths at the two radial boundaries.
A linearized multi-species collision operator is implemented in the code \cite{Esteve_PoP2015} but here only collisionless simulations are considered.
No filters in the toroidal mode number are imposed in these simulations with GYSELA, therefore  all components are allowed to develop. As shown in the following sections, the results of GYSELA are found to be in good agreement of those obtained with codes which use a linearized version of the model equations and filter out the non-zonal component. This means that, for the tests chosen in this paper, the nonlinear excitation of non-zonal components is negligible and does not sensibly modify the evolution of the zonal component.

The numerical scheme of GYSELA is based on a semi-Lagrangian method~\cite{Sonnendruecker99} (more specifically on a ``backward semi-Lagrangian approach''), which is a mix between PIC and Eulerian methods exhibiting good properties of conservation~\cite{Grandgirard06}.
In this method, the phase-space mesh grid is kept fixed in time (like in Eulerian codes) and the Vlasov equation is integrated along the trajectories (like in PIC codes) using the invariance of the distribution function along the trajectories (Liouville theorem). 
In GYSELA, the interpolation step is presently performed with cubic splines.

Like for ORB5 and GENE, the model equations of GYSELA are based on the gyrokinetic equations of Brizard and Hahm~\cite{Brizard07}. 
Then, the time evolution of the full guiding-center distribution function $F_s$ is governed for each species $s$, by the same form of equation as the one of GENE i.e. Eq.~\ref{eq:gyrokinetic-GENE} where the characteristics, i.e. the trajectories of the gyrocenters, are given by Eqs.~\ref{eq:trajectories_a-GENE} and~\ref{eq:trajectories_b-GENE}.
These 5D gyrokinetic Vlasov equations are self-consistently coupled to a 3D quasi-neutrality equation defined as:
\begin{equation}
  \label{eq:GYSELA_QN}
  - \frac{1}{n_{e_0}}\sum_s q_s \nabla_\perp\cdot\left(\frac{n_{s0}}{B\Omegas}\nabla_\perp\pot(\vecx,t)\right) + e \left(\frac{\pot-\bar\pot}{T_e}\right) = \frac{1}{n_{e_0}}\sum_s q_s ( n_s - n_{s0})
\end{equation}
where the gyrocenter density $n_s$ of species $s$ reads $n_s(\vecx,t) = \int\jacobv\dd\mu\dd\vGpar\; \tilde{F}_s(\vecx,\vecv,t)$ with $\jacobv=2\pi B_{G\|}^*/m_s$ the jacobian in velocity space. 
The equilibrium gyrocenter density $n_{s0}$ corresponds to the same expression as $n_s$ where $F_s$ is replaced by the equilibrium Maxwellian $F_{s0}$.  
The gyroaverage operator  was historically  approximated by a Pad\'e expansion but in this paper the new version based on direct integration on the gyro-circles with Hermite interpolation is used.
In GYSELA, the quasi-neutrality equation Eq.~\ref{eq:GYSELA_QN} is solved with finite differences in radial direction and Fourier projection in $\theta$ direction ($\varphi$ plays the role of a parameter). 
See Appendix A in Ref.~\cite{Grandgirard16} to see how the presence of $\bar\phi$ is overcome.

\section{Numerical simulations with adiabatic electrons}
\label{sec:ad_ele}

In this section, the results of numerical simulations of GAMs with adiabatic electrons are discussed, and compared with analytical theory. The main aim here is to perform a detailed verification and benchmark of the different gyrokinetic codes. This has the triple role of: a) understanding better the behavior of the linear GAM dynamics in different regimes; b) understanding better the behavior of the codes, which is crucial in the view of a future work where more physics is present; c) understanding better the regimes of validity of each specific analytical dispersion relation. Two main regimes are considere: one where the GAM radial size is large with respect to the ion larmor radius, and therefore the FOW effects are smaller, and one where the GAM radial structure is finer, and therefore the FOW effects are larger.

\subsection{GAMs with broad radial structure}
\label{sec:adele_broad_modes}

\subsubsection{Analytical predictions}
\label{sec:adele_anal_broad}

In the case of GAMs with broad radial structure ($k_r \rho_i \ll 1$), an analytical theory neglecting the FLR and FOW corrections can be considered as a first approximation. Although an MHD theory would be sufficient for estimating the order of magnitude of the GAM frequency, nevertheless, due to the resonances with ions, a gyrokinetic treatment of the ions is necessary for a proper estimation of the GAM frequency and damping rate.
Such a dispersion relation in the case of circular flux surfaces has been provided by F. Zonca in 1996~\cite{Zonca96} in the general electro-magnetic case, for low-frequency Alfv\'en modes, and can be adopted for GAMs as well, when neglecting diamagnetic effects, as discussed in details in Ref.~\cite{Zonca08}. No resonances of the electrons are retained. It reads:
\begin{equation}
\Lambda^2 (z) = z^2  + q^2 \omega_{ti}^2 z \Big[  F(z) - \frac{N^2(z)}{D^2(z)}    \Big] = 0
\label{eq:disp_rel_Zonca}
\end{equation}
where $z = (\omega + i \gamma)/\omega_{ti}$, $\omega_{ti}= \sqrt{2} \, v_{ti}/(q R_0)$ is the transit ion frequency, $v_{ti}=\sqrt{T_i/m_i}$, and the functions $F$, $N$ and $D$ are defined by:
\begin{eqnarray}
F(z)  & = & z (z^2 + 3/2) +  (z^4 + z^2 + 1/2) Z(z)\\
N(z)  & = & z + (1/2+z^2)Z(z) \\
D(z)  & = & \Big( \frac{1}{z}\Big) \Big( 1 + \frac{1}{\tau_e}\Big) + Z(z) 
\end{eqnarray}
where $\tau_e = T_e/T_i$ is the ratio of electron over ion temperatures, and $Z(z)$ is the plasma dispersion function:
\begin{equation}
Z(z) = \pi^{-1/2} \int_{-\infty}^{+\infty}\frac{e^{-y^2}}{(y-z)} dy
\end{equation}
Eq.~\ref{eq:disp_rel_Zonca} is the desired dispersion relation. It is in implicit form, i.e. the zeroes of the function $\Lambda(z)$ must be found in the complex plane.


For shorter wavelengths and/or larger $q$, FLR/FOW effects become more important \cite{Qiu09ppcf}, and higher order transit resonances play a more important role in the Landau damping of GAMs, in addition to the modification of their real frequency. An extension of equation (\ref{eq:disp_rel_Zonca}) to the case where FLR and FOW effects are also considered to the first order (still in circular geometry, and with adiabatic electrons), was made for general low-frequency Alfv\'en modes by F. Zonca in 1998~\cite{Zonca98}. An approximated explicit formula for the frequency and damping rates of GAMs with $\omega =l\omega_{ti}$, $l=\pm1,\pm2$ transit resonances accounted for was provided by H. Sugama in 2006~\cite{Sugama06} and 2008~\cite{Sugama08}, in the regime of moderate values of $q$:
\begin{eqnarray}
\frac{\omega}{q\omega_{ti}} & = & f_T^{1/2} \; \Big(1+ \frac{1}{q^2}\frac{f_{S1}}{f_T^2}\Big)^{1/2}\label{eq:omega-SW}\\
\frac{\gamma}{q\omega_{ti}} & = & - \frac{\sqrt{\pi}}{2}  q^3 f_T   \Big[ \exp(-x^2) \, (x^2 + 2\tau_e + 1 )  + \frac{q^2}{4} k_r^2 \rho_i^2 \exp\Big(-\frac{x^2}{4}\Big) \Big( \frac{x^4}{128} + f_{S2} x^2 + f_{S3}\Big)\Big] \label{eq:gamma-SW}
\end{eqnarray}
with $x = \omega/\omega_{ti} = \Re(z)$, $\rho_i = \sqrt{2T_i/m_i}/\Omega_i$, $f_T = 7/4 + \tau_e$ and $f_{S1} =   23/8 + 2 \tau_e + \tau_e^2/2$, $f_{S2} = (1+\tau_e)/16$, and $f_{S3} = 3/8 + 7\tau_e /16 + 5\tau_e^2/32$ (with $\Omega_i$ being the ion cyclotron frequency).
Note that, in the limit of large values of $q$ (i.e. $q>4$) and $\tau_e = 1$, the normalized frequency, Eq.~\ref{eq:omega-SW}, tends to $\omega/q\omega_{ti}\simeq f_T^{1/2}=1.66$, and $x \simeq f_T^{1/2} q = 1.66\, q$. For cases with large enough values of $q$ to satisfy $\omega \simeq 2\omega_{ti}$, the second term in equation (\ref{eq:gamma-SW}) (namely the one proportional to $k_r^2\rho_i^2$) becomes dominant since the ions with lower energy (and thus, which are present in larger number) resonate with the GAM frequency. Note that FLR corrections are not included in Eq.~\ref{eq:gamma-SW}.

The effect of elongation $e$, in a gyrokinetic treatment, has been included by Z. Gao in 2009~\cite{Gao09}, in the large aspect ratio limit, and neglecting the FLR/FOW effects. The resulting GAM frequency and damping rate, where we neglect here the effect of the radial derivative of the elongation because not taken into account in our paper, are\footnote{a typo was present in the original paper, due to a missing proper normalization of $\omega$ in the formula for the damping rate.}:
\begin{eqnarray}
\frac{\omega}{q\omega_{ti}} & = & f_T^{1/2} \; \Big(\frac{e^2+1}{2}\Big)^{-1/2} \Big(1+ \frac{e^2+1}{2}  \frac{1}{2q^2}\frac{f_{S1}}{f_T^2}\Big)\label{eq:omega-Gao} \\
\frac{\gamma}{q\omega_{ti}} & = & - \frac{\sqrt{\pi}}{2} \frac{1}{q f_T} x^6 \exp(-x^2) \label{eq:gamma-Gao}
\end{eqnarray}
Note that, for $e=1$, i.e. for circular flux surfaces, and for large values of $q$, the frequency given by Gao-2009, Eq.~\ref{eq:omega-Gao}, reduces to the one of Sugama-2008, Eq.~\ref{eq:omega-SW}. Also note that, for deriving the damping rate of Gao-2009, Eq.~\ref{eq:gamma-Gao}, one has to neglect the second of the two terms of the damping rate given by Sugama-2008, Eq.~\ref{eq:gamma-SW} (which means assuming that the values of $q$ are below a certain threshold) and at the same time assuming the limit of large values of $q$ (i.e. large values of $x$). Due to these strong approximations, we expect the formula for the damping rate of Gao-2009 to give a good qualitative comparison with the results of numerical simulations, but some divergence in the absolute values are not to be surprising.

The previous dispersion relations, namely Eq.~\ref{eq:disp_rel_Zonca} given by F. Zonca in 1996, Eqs.~\ref{eq:omega-SW} and \ref{eq:gamma-SW} given by H. Sugama in 2006 and 2008, and Eqs.~\ref{eq:omega-Gao} and \ref{eq:gamma-Gao} given by Z. Gao in 2007, are considered as a reference for the comparison with all the results of numerical simulations on GAMs with broad radial structure, shown in Sec.~\ref{sec:adele_broad_modes}. Separate dispersion relations, where higher-order FLR/FOW effects are taken into account, are discussed in Sec.~\ref{sec:adele_loc_modes} and used for comparison with the results of numerical simulations shown in the same section.

\subsubsection{Equilibrium and definition of the simulation}
\label{sec:adele_broad_eq}

For our numerical test, we choose a tokamak equilibrium with high aspect ratio ($\varepsilon=a/R=0.1$), with $R_0 = 1.3$ m, $a=0.13$ m. The equilibrium magnetic field is given by $\vecb{B} = (B_0 R_0/R) (\vecb{e}_\varphi + (r/qR_0) \vecb{e}_\theta)$. The value of the magnetic field on axis is $B_0=1.9$ T. Each simulation has a different $q$ profile, flat, and each one with different value of $q$. Flat temperature and density profiles are also always considered. The value of $\rho^* = \rho_s/a$ is chosen as $\rho^* = 1/160$ for all simulations shown in Sec.~\ref{sec:adele_broad_modes} (with $\rho_s = c_s/\Omega_i$ being the sound Larmor radius). The value of the density is irrelevant for electrostatic simulations.

We initialize a charge density perturbation with only zonal component (i.e. independent of the poloidal and toroidal angle), and generating a scalar potential with a sine dependence on the radius, of the form $\phi(\rho,t=0) = \sin(k_r a \rho)$, with $k_r = 2\pi/a$ (where $\rho=r/a$ is the normalized minor radius, with values in [0,1]). In GYSELA, the initial perturbation of the distribution function is chosen such that $\phi(\rho,t=0) = 1-\cos(k_r a \rho)$. 
This choice has been preferred because the radial profile of the zonal component stays more stable in time than for the case $\phi(\rho,t=0) = \sin(k_r a \rho)$, leading to a mean $k_r$ value in time closer to the initial one. 
This is particularly true for large values of $k_r$ as those explored in section \ref{sec:adele_loc_modes}.
One explanation could be that in the case of $1-\cos$ profile the gradients are flatter at radial boundaries so that boundary conditions seem to have less impact.
Anyway, this raise the delicate point of confronting global nonlinear code results with linear theory results which are based on local approximation. With this choice of $\rho^*$ and $k_r$, we obtain a relatively low value of $k_r \rho_i$, namely $k_r \rho_i=0.055$  which corresponds to a regime where ion FOW effects are relatively small, for moderate values of $q$. The perturbation is let evolve in a linear electrostatic simulation with adiabatic electrons. 
GAMs oscillations are observed, and we measure the scalar potential, and calculate frequency and damping rate.


\subsubsection{Dependence on the safety factor}
\label{sec:adele_broad_q}

\begin{figure}[b!]
\begin{center}
\includegraphics[width=0.49\textwidth]{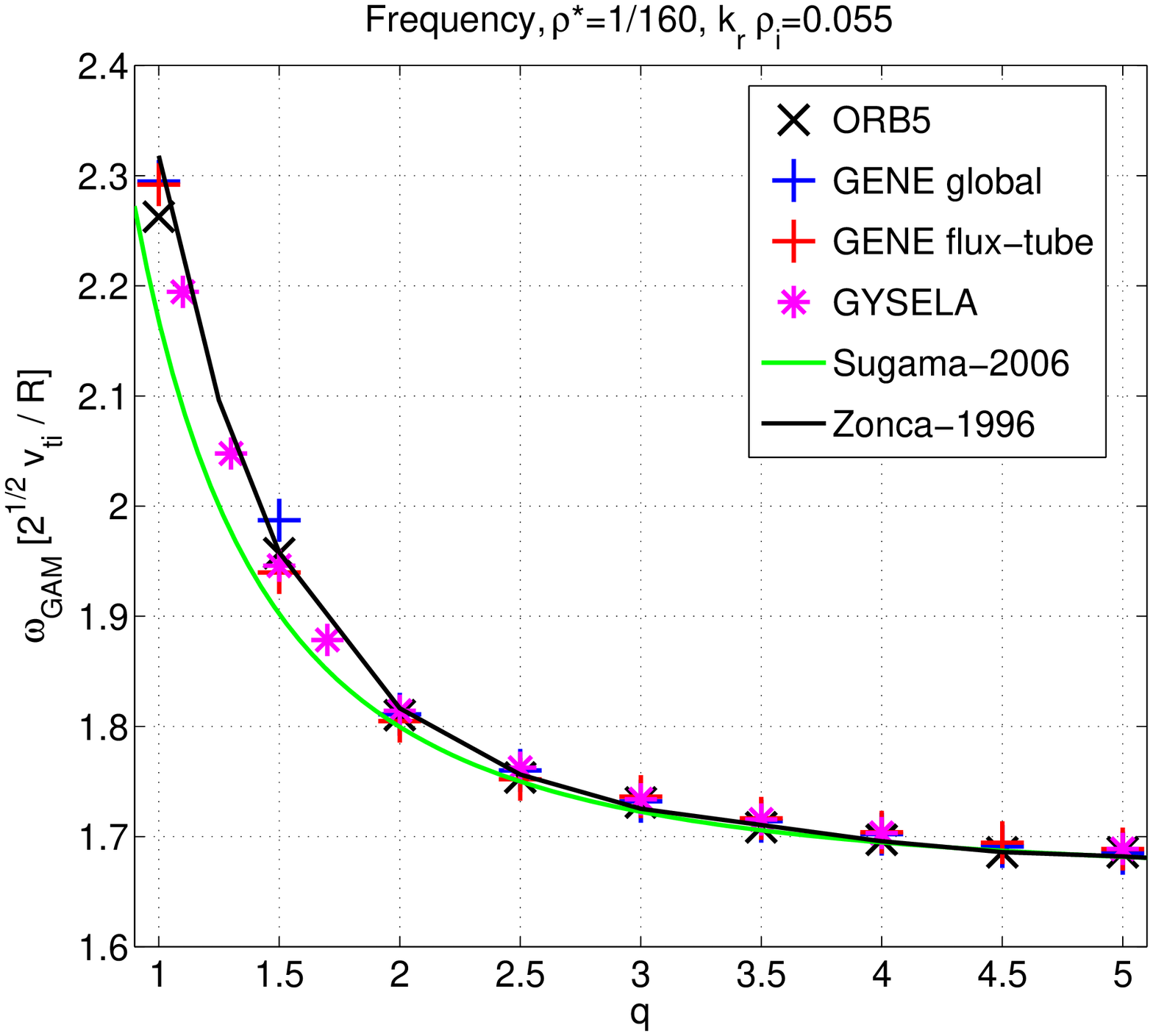}
\includegraphics[width=0.49\textwidth]{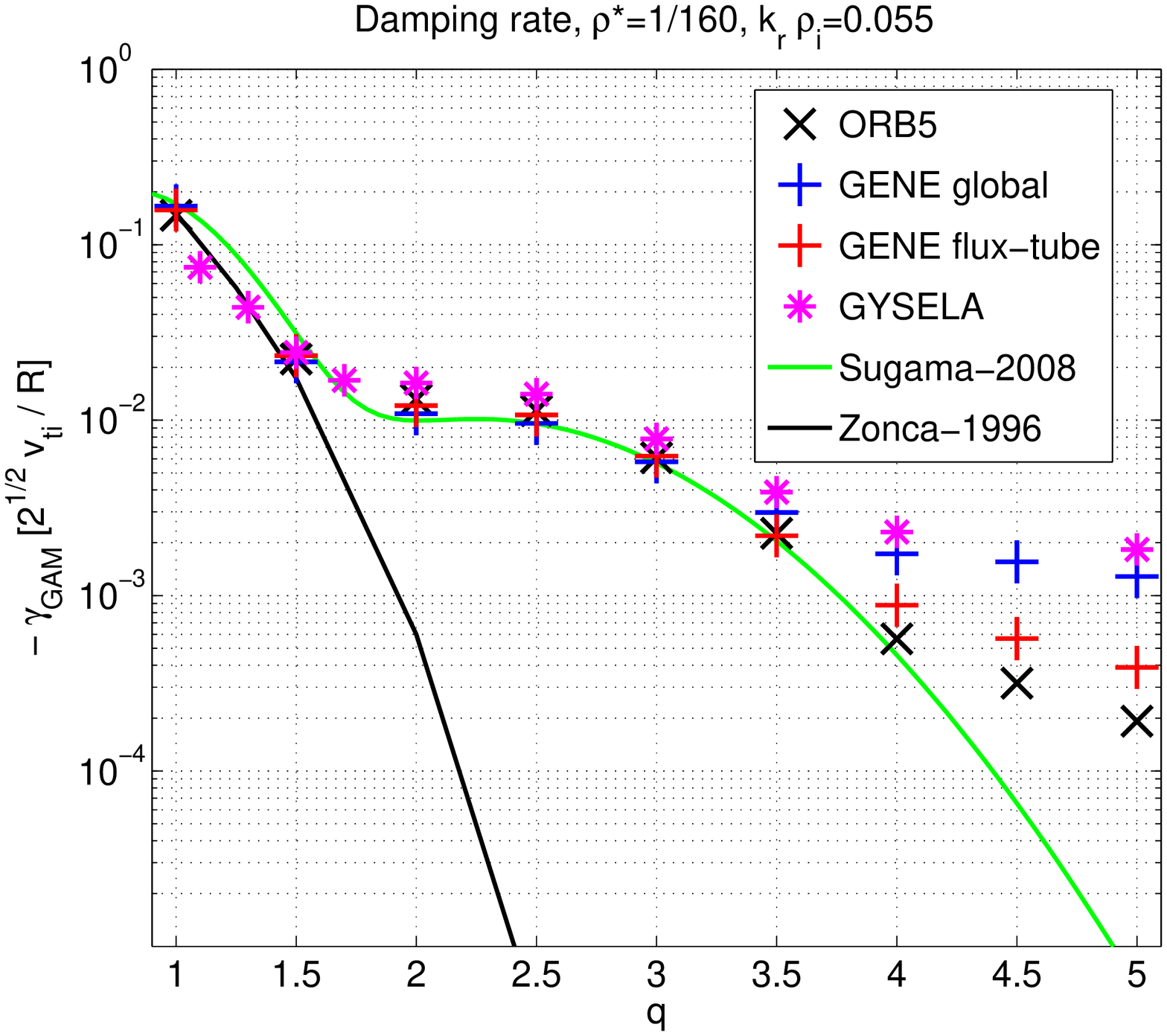}
\vskip -1em
\caption{\label{fig:omegagamma_q} Frequency (on the left) and damping rate (on the right), measured with ORB5 (in black Xs), GENE global (blue crosses), GENE flux-tube (red crosses), and GYSELA (magenta stars). The results of the explicit analytical formulas of Sugama-2006 (Eq.~\ref{eq:omega-SW}) and Sugama-2008 (Eq.~\ref{eq:gamma-SW}) are also shown in green, and of the dispersion relation of Zonca-1996 (Eq.~\ref{eq:disp_rel_Zonca}) in black.}
\end{center} 
\end{figure}

For the simulations shown in this section, we have considered an analytical equilibrium with concentric circular flux surfaces. The radial electric field is measured at the radial position of its peak, namely at mid-radius, $\rho=0.5$. It is observed to oscillate in time, and be damped due to Landau damping.
The frequency is measured for different simulations  with different value of q, obtained with ORB5, GENE and GYSELA, and it is found to scale correctly with the theoretical dispersion relation Zonca-1996 (Eq.~\ref{eq:disp_rel_Zonca}), and the explicit formula Sugama-2006 (Eq.~\ref{eq:omega-SW}), as shown in Fig.~\ref{fig:omegagamma_q}. In particular, note that the value of the frequency tends to $\omega_{q\rightarrow\infty}/q\omega_{ti} = 1.66$ for large values of $q$, as discussed in Sec.~\ref{sec:adele_anal_broad} after Eq.~\ref{eq:omega-SW}. Some minor differences are found at low values of $q$ in the results of the dispersion relations (due to the hypothesis of large $q$ considered by Sugama for the calculation of the explicit formula).

The dependence of the damping rate on $q$ has also been studied, for the same simulations (see Fig.~\ref{fig:omegagamma_q}). All codes match well with the analytical predictions of Zonca-1996 (Eq.~\ref{eq:disp_rel_Zonca})  at low values of $q$ ($q\le 1.5$), where the FOW effects are negligible. At larger values of $q$ ($q > 1.5$), the FOW effects included at the first order in the explicit formula Sugama-2008 (Eq.~\ref{eq:gamma-SW}) are shown to be dominant. All codes fit well with Sugama-2008 for values of $q$ smaller than 3.5. At even larger values of $q$ ($q\ge 3.5$) the higher-order FOW effects become dominant, and deviations from the formula Sugama-2008 are observed. This regime is studied more in details in Sec.~\ref{sec:adele_loc_modes}.
The difference at large q between the flux-tube version of GENE (which agrees perfectly with ORB5)  and the global version of GENE  is due to the fact that the $k_r$ used for global GENE runs was slightly larger. For this value of $\rho^*$, the choice of $k_r \rho_i$ = 0.055 requires to simulate the entire domain in minor radius, while simulations of global GENE accounted only for 98\% of it. This affects only the high q, i.e. when the damping is very small and the relative effect of $k_r$ is large. Values of damping rates larger than ORB5 at large values of q are also observed with GYSELA, probably because the value of $k_r$ has been observed to evolve in time towards values which are a bit larger than at the initial time of the simulations with GYSELA, and this increases the averaged damping rate.

\subsubsection{Dependence on the elongation}
\label{sec:adele_broad_elo}


\begin{figure}[b!]
\begin{center}
\includegraphics[width=0.49\textwidth]{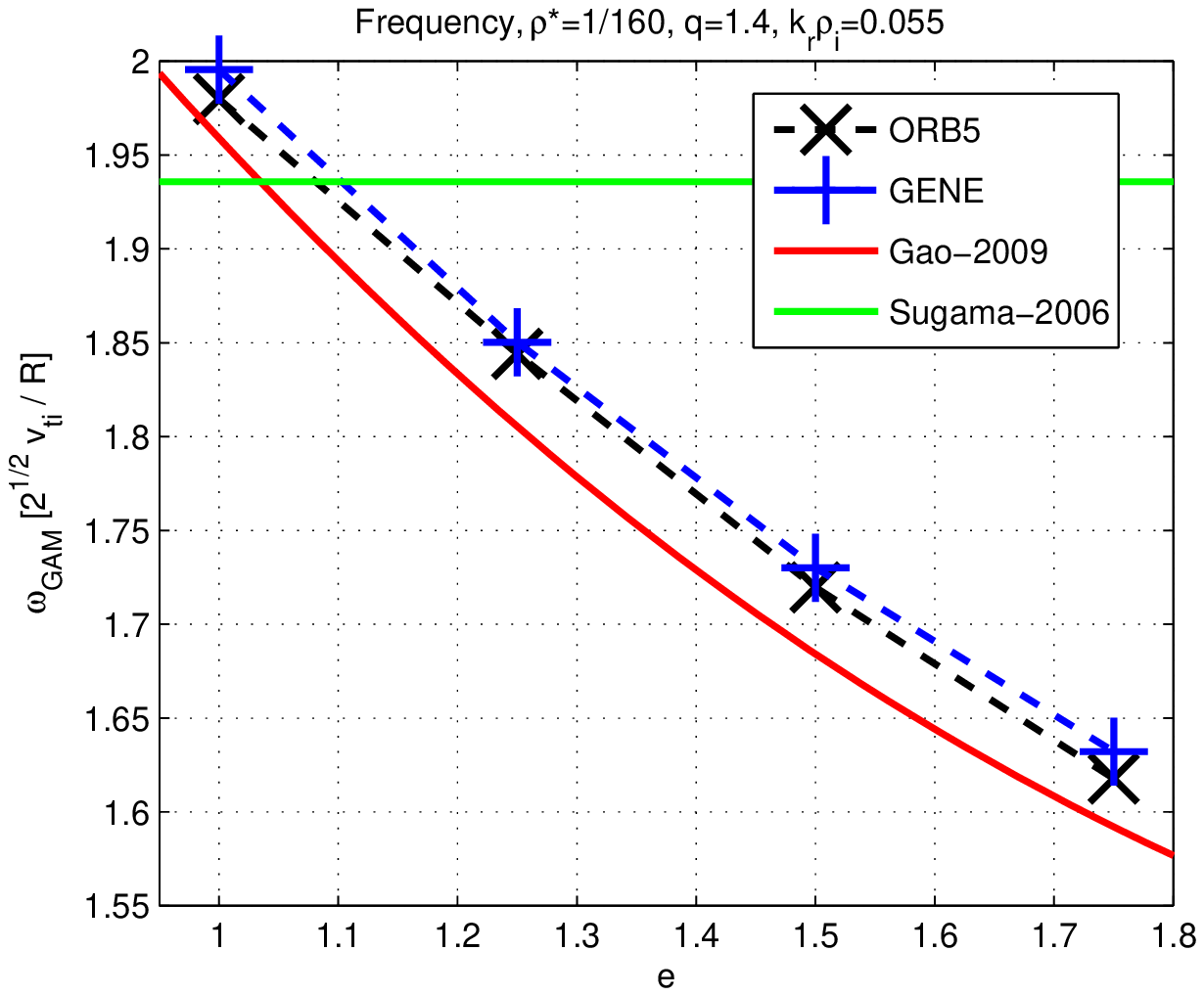}
\includegraphics[width=0.49\textwidth]{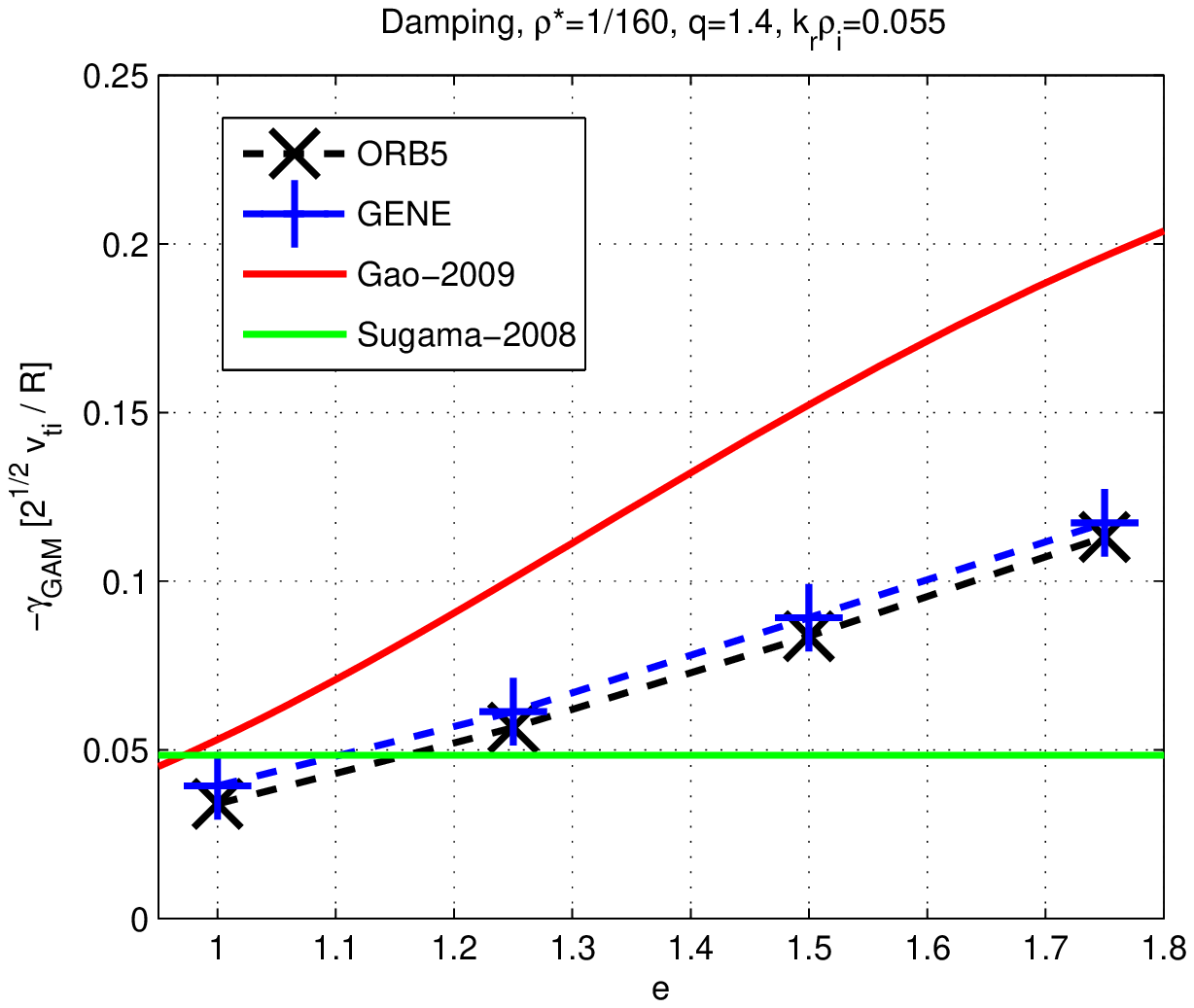}
\vskip -1em
\caption{\label{fig:omegagamma_e} Frequency (on the left) and damping rate (on the right) of the radial electric field, measured with ORB5 (in black Xs), and GENE global (blue crosses). The analytical formulas of Sugama-2006 and Sugama-2008 are also shown (in green), and of Gao-2009 in red.}
\end{center} 
\end{figure}

The dependence on the elongation has been studied by loading magnetic equilibria with different elongation (and no triangularity) calculated with the CHEASE code~\cite{Luetjens96}. These simulations have been performed with ORB5 and GENE. The safety factor has been chosen with $q=1.4$, and we have varied the elongation from $e=1$ (circular flux surfaces) to $e=1.75$ (elongated plasmas). The results are shown in Fig.~\ref{fig:omegagamma_e}.

The frequency measured with the two codes has been found to fit very well, falling within the error bars for the whole scan. The fit with the analytical prediction of Gao-2009 is also very good (with a maximum of 3\% of difference not depending on the elongation), showing the correct decrease of the frequency with the increasing  elongation.

Regarding the damping rate, a very good matching of the two codes has also been found, showing an increase of the damping rate with the elongation, as predicted by the analytical theory. A quantitative fit of the damping rate with the analytical theory has been found worse than for the frequency. This is probably due to the fact that the dependence of the dispersion relation on the safety factor $q$, the FOW effects, proportional to $k_r^2 \rho_i^2$, and the elongation $e$ at the same time, forces some strong approximations to be taken when deriving an explicit analytical formula, as discussed in Sec.~\ref{sec:adele_anal_broad}. Therefore, the analytical derivation is based on some assumptions, like the assumption of negligible FOW effects, and of large values of $q$ at the same time, which is most likely at the origin for the divergences with the results of the numerical simulations. Note that this difference, up to 40\%, of the result of the numerical simulations with respect to the analytical theory is of 
the same order of magnitude of what found also in the previous scan, shown in Sec.~\ref{sec:adele_broad_q}, for the value of $q=1.4$. This confirms that the quantitative analytical prediction of the damping rate is very challenging, due to the many approximations needed in deriving explicit formulas.

\subsection{GAMs with arbitrary radial structure}
\label{sec:adele_loc_modes}

In this section, we want to investigate the linear collisionless dynamics of GAMs in a regime where the FOW effects are more important, therefore we push towards higher values of $k_r \rho_i$, corresponding to GAMs with finer radial structure with respect to the ones considered in Sec.~\ref{sec:adele_broad_modes}. We neglect here the effect of the elongation, and we still consider only the results obtained by the analytical theory and numerical simulations with adiabatic electrons.

\subsubsection{Analytical predictions}
\label{sec:adele_anal_local}

As $k_r\rho_i q^2$ further increases, higher and higher order transit resonances must be taken into account to properly get more accurate GAM damping rates, as it was first shown in Ref.~\cite{Xu08prl}, and discussed in details by Z. Qiu in 2009~\cite{Qiu09ppcf}. The collisionless damping of GAMs for $k_r\rho_iq^2\rightarrow\infty$ was derived in Ref.~\cite{Zonca08} with all the transit resonances FOW and FLR properly accounted for, and the dispersion was later extended to relatively smaller $q$ parameter region in Ref.~\cite{Qiu09ppcf} to compare with numerical simulations \cite{Xu08prl}. The dispersion relation of Qiu-2009 was calculated in the limit of large values of $q$ and moderate values of $k_r\rho_i$, i.e. $1/q^2 \ll k_r \rho_i \ll 1$. It reads:
\begin{eqnarray}
\frac{\omega}{q\omega_{ti}} & = & f_T^{1/2} \; \Big[ 1 + \frac{1}{2q^2} \frac{f_{S1}}{f_T^2}  + \frac{\hat{k}^2}{4} \Big( - \frac{f_{Q1}}{f_T} + \frac{f_{Q2}}{f_T^2}\Big) \Big] \label{eq:omega-Qiu}\\
\frac{\gamma}{q\omega_{ti}} & = & - \frac{\sqrt{2}}{\hat{k}^2}\frac{\exp(-\hat\omega/\hat{k})}{\hat\omega^5}\Big[ \hat\omega^4 +\frac{f_{Q1}}{2} \hat\omega^2 \hat{k}^2 -f_{Q2}\hat{k}^2 -2 f_{S1}\frac{1}{q^2}\Big] \cdot \nonumber \\ 
& & \cdot \Big[ \hat\omega^2  + \tau_e \hat\omega \hat{k} + f_{Q3}\hat{k}^2 - \hat\omega^2\hat{k}^2 - \frac{\hat\omega^3}{8} \frac{1}{q^2 \hat{k}^3} + \frac{\hat\omega^4}{24} \frac{1}{q^2 \hat{k}^4}\Big] \label{eq:gamma-Qiu}
\end{eqnarray}
where $\hat\omega = \omega/q\omega_{ti} = x/q$, $\hat{k} = k_r\rho_i$,  $f_{Q1} = 31/16 + 9 \tau_e/4 + \tau_e^2 $, $f_{Q2} = 747/32 + 481 \tau_e/32 + 35\tau_e^2/8 + \tau_e^3/2 $, and $f_{Q3} = \tau_e^2 + 5 \tau_e/4 + 1$. Note that, differently from the analytical predictions described in Sec.~\ref{sec:adele_broad_modes}, the GAM frequency has a dependence on $k_r \rho_i$. The $k_r\rho_i$ in Eq.~\ref{eq:gamma-Qiu} comes from both FLR ($J^2_0(k_r\rho_i)$) and also FOW ($J_p(k_r\rho_d)$, with p being integers, and $k_r\rho_d\simeq k_r\rho_i q$ for circulating particles.
In the limit of large values of $q$ for a fixed $\hat{k}$, then $\hat\omega$ tends to a constant with value $\hat\omega_{q\rightarrow\infty}\simeq f_T^{1/2} $. In this limit, the last term in the first squared bracket of the formula for the damping rate, Eq.~\ref{eq:gamma-Qiu}, and the last two terms in the second squared bracket of the same formula, can be neglected, and the GAM damping rate tends to a constant value. Note that Eqs.~\ref{eq:omega-Qiu}, \ref{eq:gamma-Qiu} can be used for both short/long wavelength regimes. E.g., in long wavelength limit, with $k_r\rho_i\ll1$, FOW effects are still important if $q$ is large. So, in general, when we say $q$ is large or small, it is not compared to 1, but to $\sqrt{1/k_r\rho_i}$.

Finally, an analytic dispersion relation where the effects of the non-circular geometry are also included has been derived by Z. Gao in 2010~\cite{Gao10pop}. We report here the formulas for $\omega$ and $\gamma$ where no radial derivative of the elongation is considered, for concentric flux surfaces (i.e. with no Shafranov shift), and neglecting the effects of finite inverse aspect ratio. It reads:
\begin{eqnarray}
 \frac{\omega}{q\omega_{ti}} & = & f_T^{1/2} \Big(\frac{e^2+1}{2}\Big)^{-1/2}\; \Big[ 1 + \frac{e^2+1}{2} \frac{1}{2q^2} \frac{f_{S1}}{f_T^2}  + \frac{\hat{k}^2}{4 e^2} Q \Big] \\
  \frac{\gamma}{q\omega_{ti}} & = & - \frac{\sqrt{2} f_T^{1/2}}{\hat{k}^2} e^2 \Big( \frac{e^2+1}{2} \Big)^{-3/2} \exp\Big( - \frac{f_T^{1/2}}{\hat{k}} \Big( \frac{e^2+1}{2e^2} \Big)^{-1/2} \Big)
\end{eqnarray}
where
\begin{displaymath}
Q =  \frac{f_{Q2}}{f_T^2} \Big(\frac{e^2+1}{2}\Big)-   \frac{f_{G1}}{f_T}e^2 -  \frac{f_{G2}}{f_T} + \frac{f_{G3}(e^4 + 1) + f_{G4}e^2}{e^2 (e^2 + 1)}
\end{displaymath}
and $f_{G1} = (13-2\tau_e - 4 \tau_e^2)/16$, $f_{G2} = (39 + 50\tau_e +20 \tau_e^2)/16$, $f_{G3} = (9 + 4\tau_e)/16$ and $f_{G4} = (6 - 8\tau_e)/16$. In the case of no elongation, $e=1$, the dispersion relation of Gao-2010 reads:
\begin{eqnarray}
 \frac{\omega}{q\omega_{ti}} & = & f_T^{1/2} \Big( 1 + \frac{1}{2q^2} \frac{f_{S1}}{f_T^2}  + \frac{\hat{k}^2}{4} \Big(  -\frac{f_{Q1}}{f_T} + \frac{f_{Q2}}{f_T^2} \Big) \Big) \label{eq:omega-Gao-2010}\\
  \frac{\gamma}{q\omega_{ti}} & = & - \frac{\sqrt{2} f_T^{1/2}}{\hat{k}^2}  \exp\Big( - \frac{f_T^{1/2}}{\hat{k}}  \Big) \label{eq:gamma-Gao-2010}
\end{eqnarray}
Note that the frequency reduces exactly to the one of Qiu-2009. The damping rate can be derived as an approximation of the one of Qiu-2009, in the large-$q$ regime, and when considering only the largest terms in the squared parenthesis of Eq.~\ref{eq:gamma-Qiu}, i.e. respectively $\hat\omega^4$ and $\hat\omega^2$. Some differences are therefore expected for the damping rates of Qiu-2009 and Gao-2010.

\subsubsection{Equilibrium and definition of the simulation}
\label{sec:adele_loc_eq}

 We choose a tokamak equilibrium with circular flux surfaces and high aspect ratio ($\varepsilon=a/R=0.1$), with $R_0 = 1.3 m$, $a=0.13 m$. Each simulation has a different q profile, flat, and each one with different value of q. Flat temperature and density profiles are considered. Different values of $\rho^*$ are considered (and $T_i=T_e$). The value of the density is irrelevant for electrostatic simulations.
The initialization is done in a similar way as described in Sec.~\ref{sec:adele_broad_eq}, but we initialize here different simulations with different value of $k_r$.


\subsubsection{Dependence on the safety factor}
\label{sec:adele_loc_q}

\begin{figure}[t!]
\begin{center}
\includegraphics[width=0.49\textwidth]{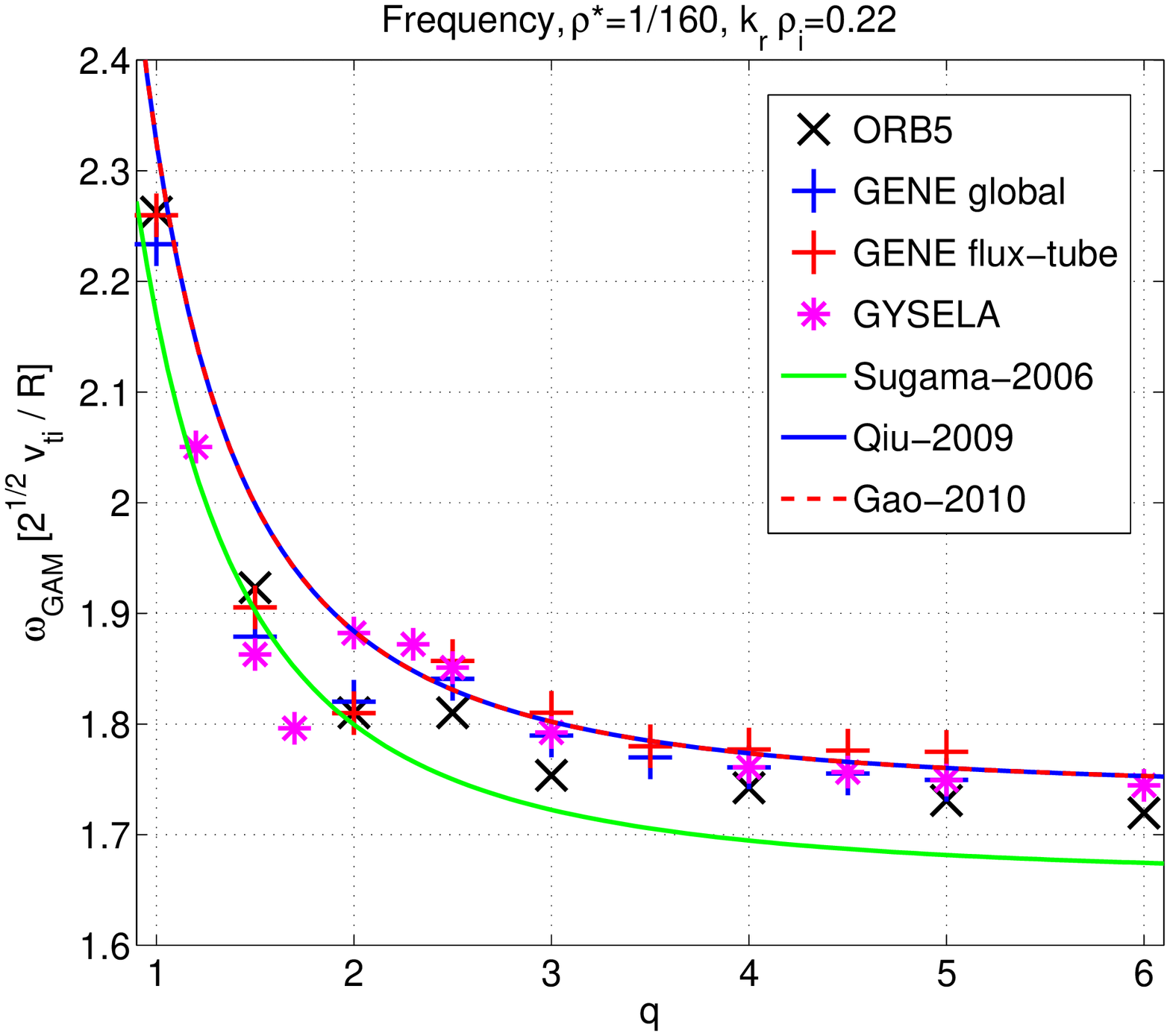}
\includegraphics[width=0.49\textwidth]{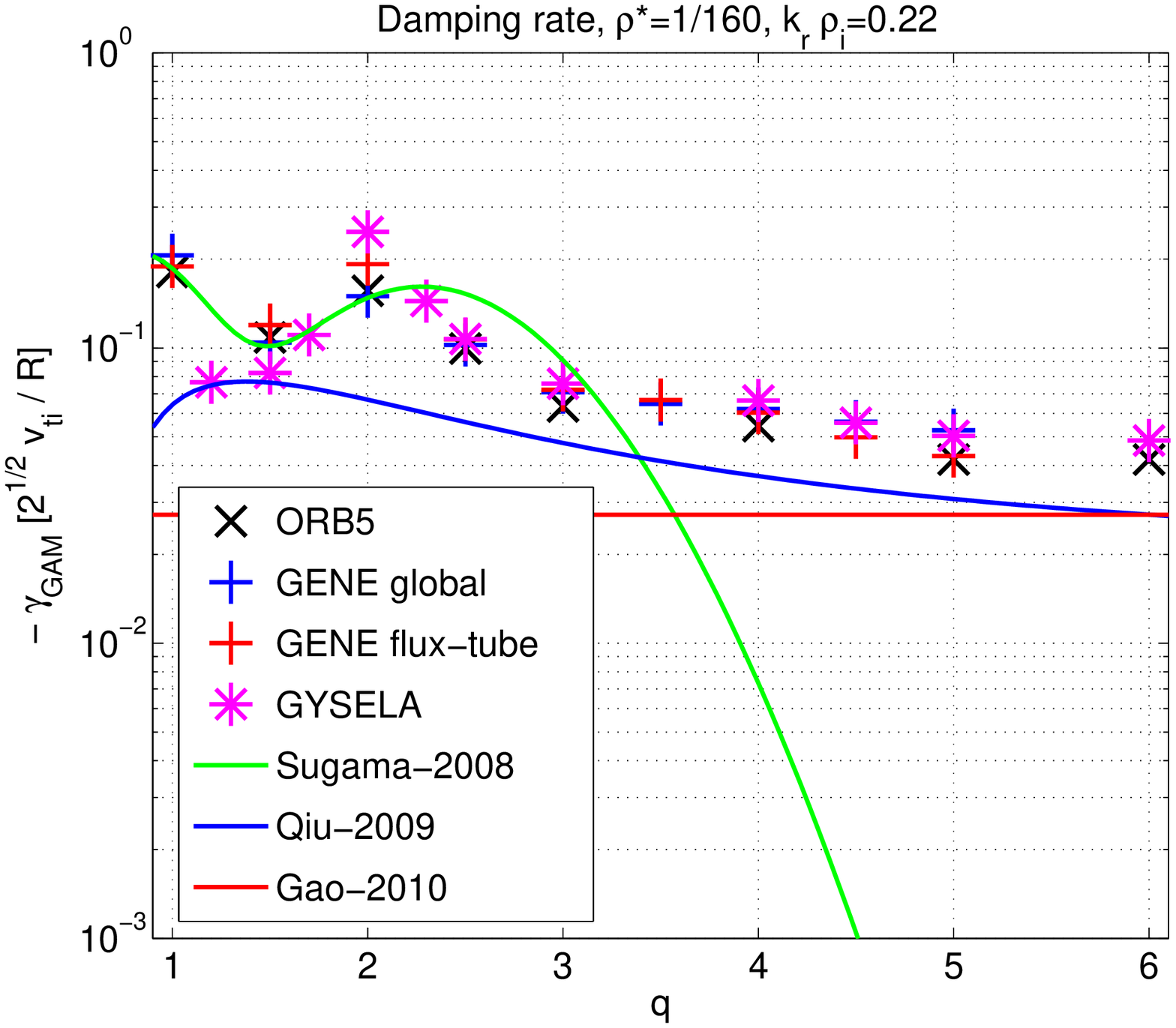}
\caption{\label{fig:gamma_q} Frequency and damping rate of the radial electric field vs q, measured with ORB5 (black Xs), GENE global (blue crosses) and GENE local (red crosses) and compared with analytical theories of Sugama-2006, Sugama-2008, Qiu-2009 and Gao-2010.}
\end{center} 
\end{figure}

A scan with $q$ has been repeated here with ORB5, GYSELA and GENE, similarly to the one reported in Sec.~\ref{sec:adele_broad_q}. The value of $\rho^*$ here has been chosen as in Sec.~\ref{sec:adele_broad_modes}, i.e. $\rho^* = 1/160$.

Frequency and damping rates of GAMs depend on the safety factor q. To the lowest order in $k_r \rho_i$, the frequency is well described by the limit of $k_r \rho_i \rightarrow 0$, and the FOW effects provide corrections which do not modify the order of magnitude of the frequency. All codes seem to follow the analytical prediction obtained without FOW effects at low values of q, whereas there is a change in trend occurring around q=2, where all codes start following the analytical predictions where the FOW effects are included in the frequency.

For the damping rates, the trend of the dependence on q is well described by the limit of small $k_r \rho_i$, where first order corrections (i.e. accounting for the 2nd harmonic resonance $v_\parallel = q R \omega_{GAM}/2$ of the passing ions), only at small q ($q < 3$).
At larger values of q ($q > 3-4$), higher order corrections (i.e. accounting for the 4nd harmonic resonance and higher) are necessary for estimating analytically the GAM damping rate. Note that, in the limit of large values of $q$, the damping rate tends to a constant, as predicted by the analytical theory, Eq.~\ref{eq:gamma-Qiu}.

As a result of this verification test, a good agreement in the scalings measured with ORB5, GYSELA and GENE (both local and global) and with the theoretical prediction of the analytical theory is found, both for the frequency and the damping rate.

\subsubsection{Dependence on the radial wave number}
\label{sec:adele_loc_kr}

The dependence of the frequency and the damping rate on the radial wave number is discussed here. As shown in the previous section, the frequency is well described by the limit of zero FOW to the lowest order, and the corrections of the FOW effects to the value of the frequency provide some modifications, up to 10\%.
The damping rate dependence on $k_r \rho_i$, on the other hand, must be considered to orders higher than the first, when $k_r \rho_i > 0.1$, if realistic values of $q$ are considered as measured in tokamaks ($q>4$). Good agreement of ORB5, GYSELA and GENE (both local and global) and the analytical theory is observed for the frequency at low values of $k_r \rho_i$, while at higher values of $k_r \rho_i$, the numerical codes predict slightly lower frequencies with respect to the analytical theory.  The origin of this discrepancy is thought to be the breaking of the regime of validity of the analytical predictions, derived with the hypothesis of moderate values of $k_r \rho_i$.

\begin{figure}[t!]
\begin{center}
\includegraphics[width=0.49\textwidth]{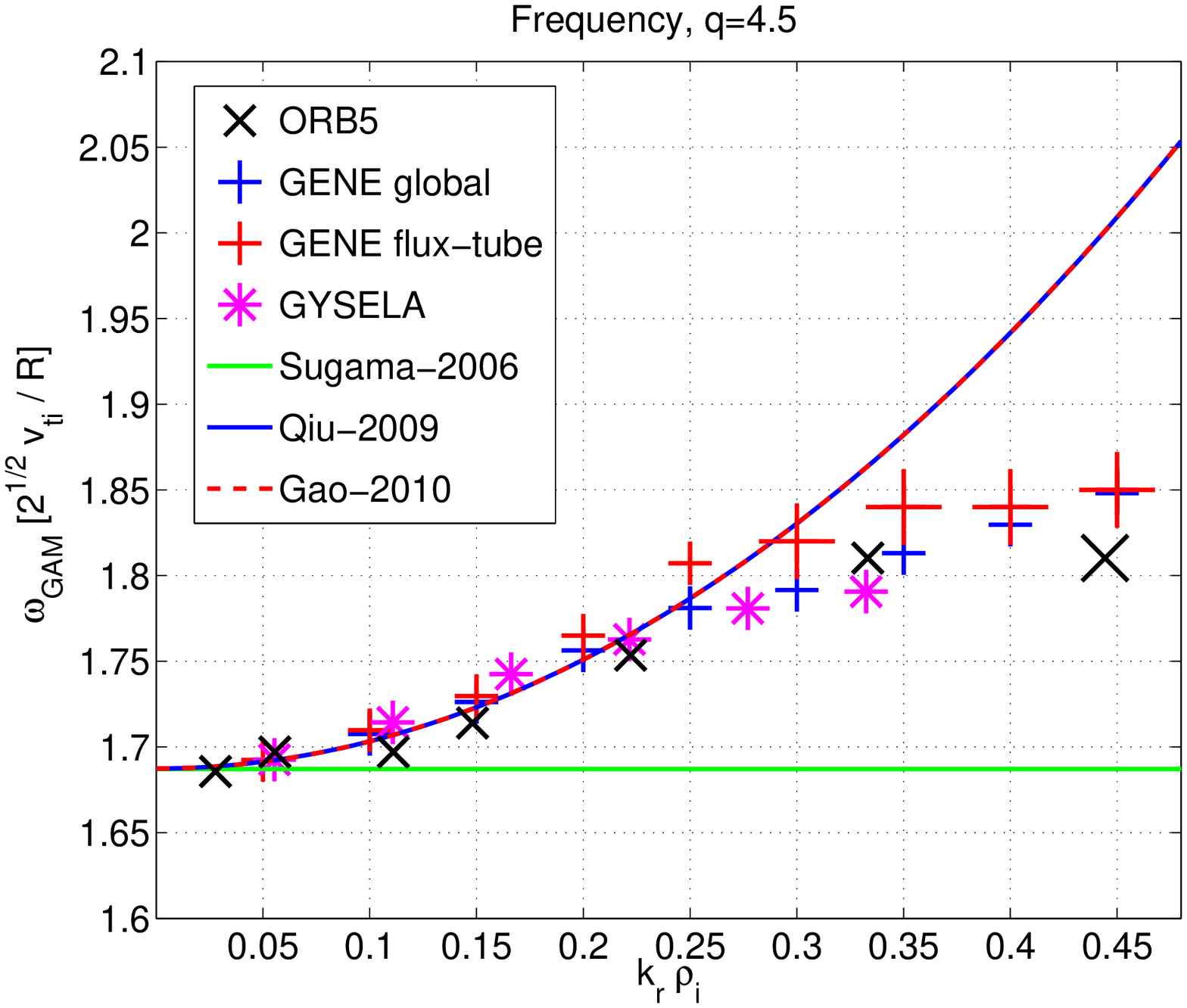}
\includegraphics[width=0.49\textwidth]{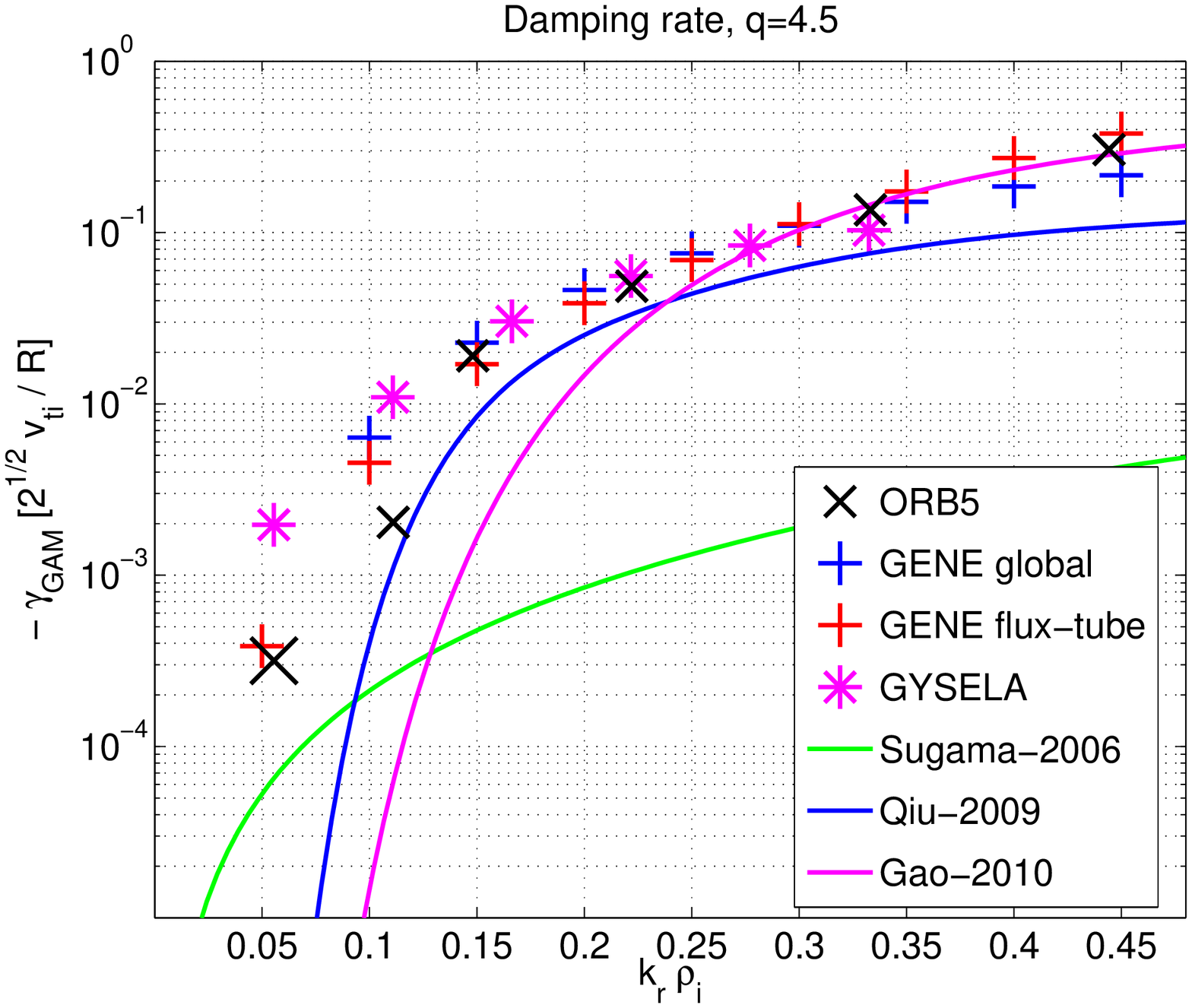}
\vskip -1em
\caption{\label{fig:omegagamma_k} Frequency and damping rate of the radial electric field vs $k_r \rho_i$, measured with ORB5 (black Xs), GENE global (blue crosses), GENE flux-tube (red crosses), and GYSELA (magenta stars), and compared with analytical theories of Sugama-2006, Sugama-2008, Qiu-2009, and Gao-2010.}
\end{center} 
\end{figure}

A very good agreement of all codes is observed for the damping rate, except at low values of $k_r \rho_i$. This is the regime where the damping rates are very small and therefore very difficult to measure, in some cases hidden below the noise (especially for PIC codes). In particular, with PIC codes the cases at very low damping rate require a very high resolution (i.e. a large number of markers) in order for the signal to overcome the statistical error. Therefore, for very low values of the damping rate, the measured numerical value is less trustable, and the error bar becomes bigger. The comparison of the gyrokinetic simulations with the three different analytical formulas of Sugama-2008, Qiu-2009 and Gao-2010 shows that the damping rate is better approximated by Sugama-2008 at very low values of $k_r \rho_i$ (although this formula still underestimates the damping rate at this large values of $q$), by Qiu-2009 at intermediate values, and by Gao-2010 at large values (see Fig.~\ref{fig:omegagamma_k}).

\section{Numerical simulations with kinetic electrons}
\label{sec:kin_ele}

\subsection{Effect of the finite electron mass, for radially broad modes}
\label{sec:kinele_me}

In this section, the same equilibrium as in Sec.~\ref{sec:adele_broad_eq} is adopted. The flux surfaces are circular, and the safety factor profile is flat, with $q=3.5$. We initialize a scalar potential perturbation with only zonal component, and with a sine dependence on the radius, of the form $\phi(\rho,t=0) = \sin(k_r a \rho)$, with $k_r = 2\pi/a$ (corresponding to a relatively low value of $k_r \rho_i$).
The perturbation is let evolve in a linear electrostatic simulation with kinetic electrons.  Our simulations have a spatial grid of (s,$\theta$,$\phi$) = 64x64x4 and a time step of 2 $\Omega_i^{-1}$, with $10^8$ markers. The length of the simulations is $4\cdot 10^4 \, \Omega_i^{-1}$, corresponding to 20000 time steps.

\begin{figure}[t!]
\begin{center}
\includegraphics[width=0.48\textwidth]{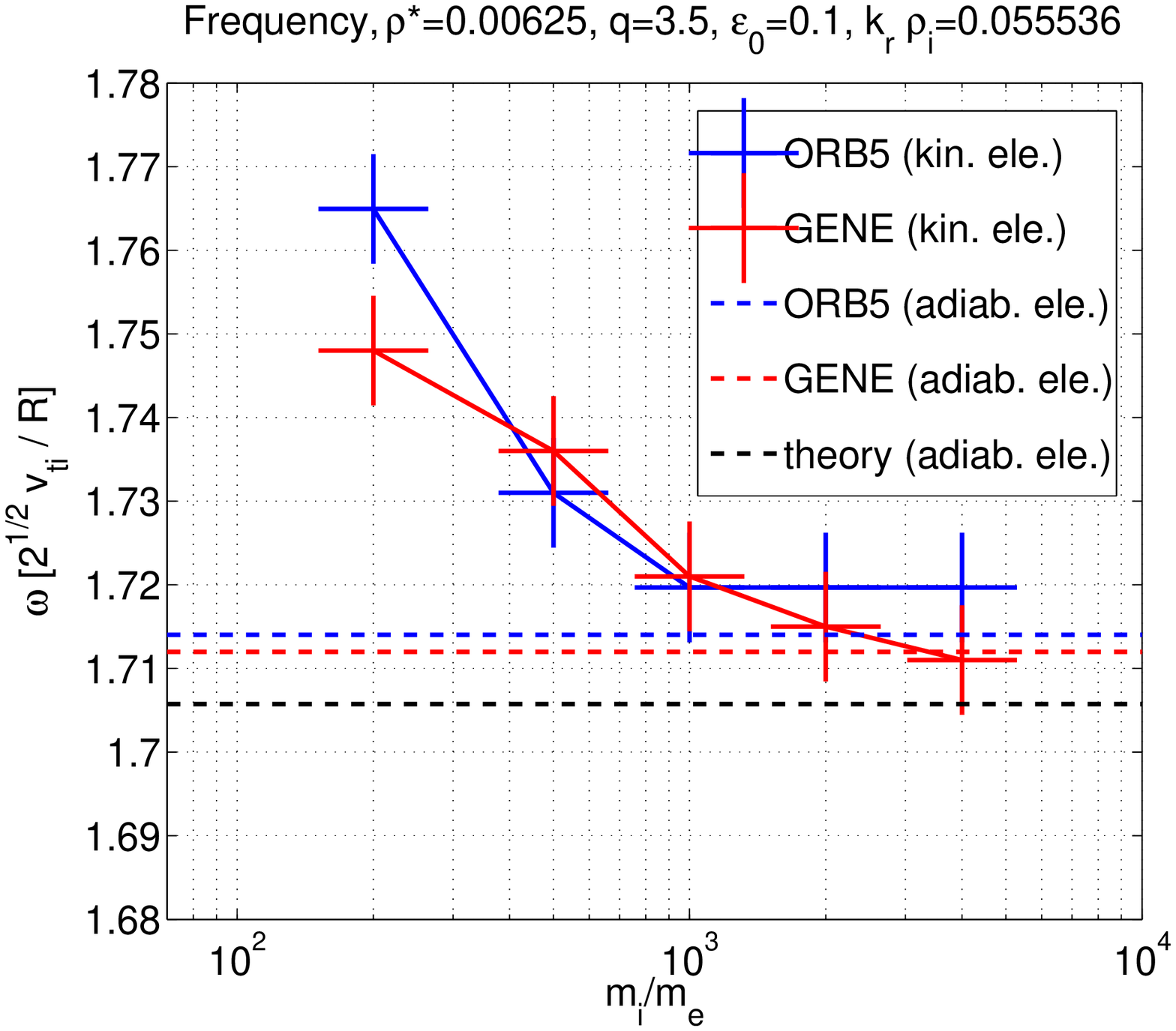}
\includegraphics[width=0.48\textwidth]{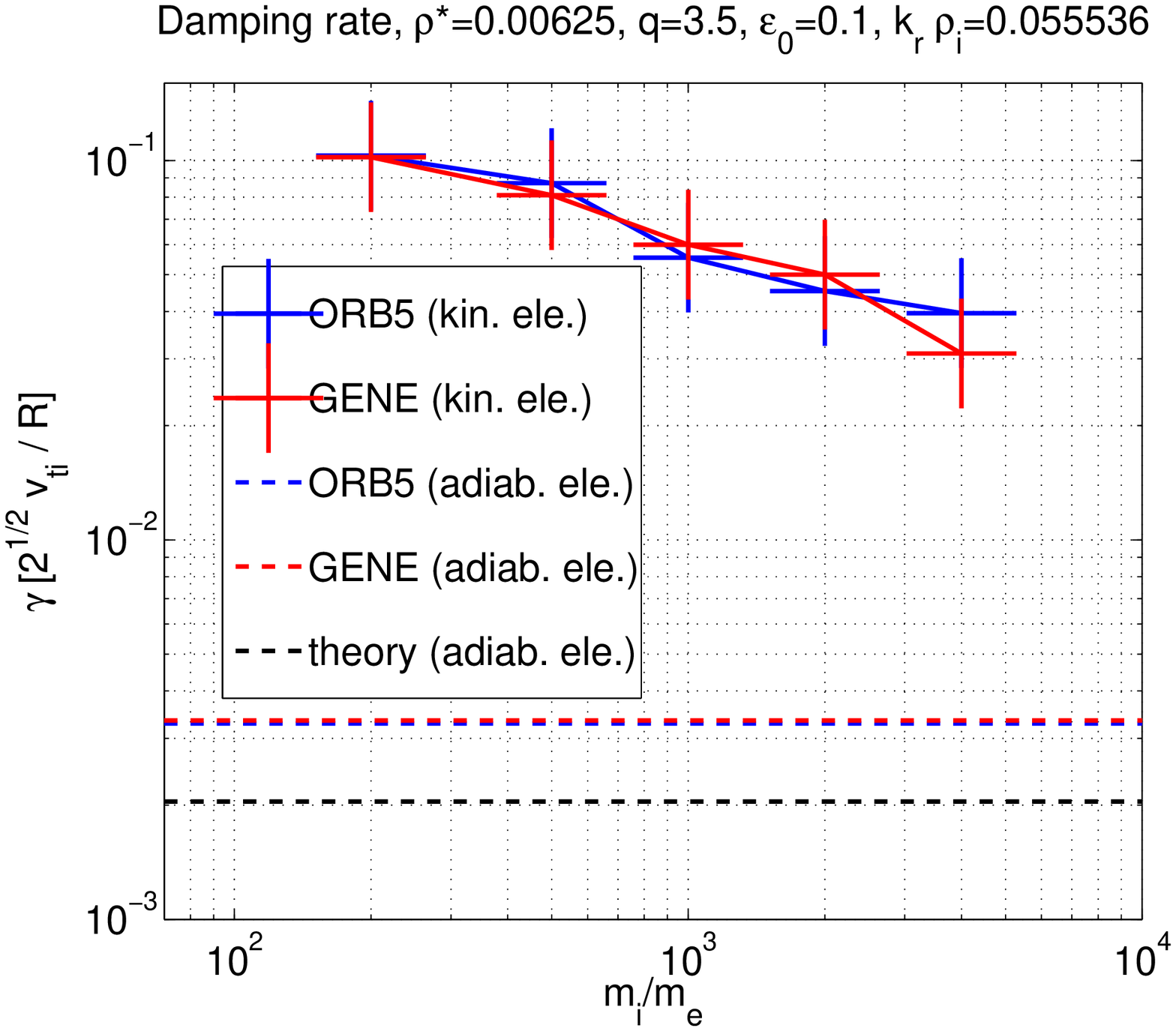}
\vskip -1em
\caption{\label{fig:FKE-omegagamma_me} Dependence of the frequency (left) and damping rate (right) on the ion/electron mass ratio, measured with ORB5 (blue crosses) and GENE (red crosses). The values obtained with adiabatic electrons are depicted as dashed horizontal lines. Circular flux surfaces are considered here.}
\end{center} 
\end{figure}

The dependence of the frequency and damping rate on the ion/electron mass ratio is depicted in Fig.~\ref{fig:FKE-omegagamma_me}, for simulations performed with ORB5 and the global version of GENE. We can see that for the frequency, a convergence towards the values of the adiabatic electrons is observed very soon for increasing $m_i/m_e$, whereas for the damping rate, the convergence is not observed. For realistic values of $m_i/m_e$ in deuterium plasmas, the measured damping rate is more than 10 times larger than the value given by the adiabatic electrons, for the chosen value of the safety factor ($q=$ 3.5). A good agreement of the two codes is found for both frequencies (giving results within 2\% of difference) and damping rates (within 25\% of difference at large mass ratios). Such a difference in the damping rate measured in simulations with kinetic electrons and with adiabatic electrons is due to the effect of the resonance with the bounce motion of barely trapped electrons~\cite{Zhang10}.

\vskip 2em

\section{Summary and conclusions}
\label{sec:conclusions}

Zonal (i.e. axisymmetric) poloidal flows, corresponding to zonal radial electric fields, are known to develop in tokamak plasmas, as the result of nonlinear interaction with turbulence. They appear in the form of zero-frequency zonal flows (ZFZF)~\cite{Hasegawa79,Rosenbluth98,Diamond05} and oscillating zonal flows, named geodesic acoustic modes (GAM)~\cite{Winsor68,Zonca08,Conway11}. Their different behavior in time results in a different efficiency in the turbulence regulation. Both ZFZFs and GAMs are crucial to be understood (linearly and then nonlinearly) for a theoretical characterization of a turbulent plasma. In this paper, we have focused on the linear collisionless dynamics of GAMs. 

The linear collisionless theory of GAMs has been developed in different regimes and several numerical investigations have been performed and compared with the theory in the past. Many gyrokinetic codes have also been developed for the study of the nonlinear interaction of turbulence and zonal structures.  Nevertheless, no comprehensive linear verification and benchmark effort has been done, to test multiple gyrokinetic codes comparing them with each other and the different analytical theories derived in different limits.

In this paper, we have selected a list of tests which serve for investigating the behaviour of some of the most known gyrokinetic codes in the magnetic-confinement-fusion turbulence community, especially in comparison with each other or with analytical theory. The choice of the codes has been made in order to give an approximative representation of the big variety of turbulence codes existing in our community.
The chosen codes have been ORB5~\cite{Jolliet07,Bottino11,Bottino15JPP}, GENE~\cite{Jenko00,Goerler11}, and GYSELA~\cite{Grandgirard06,Grandgirard16}. These codes are based on the same basic gyrokinetic formalism for the treatment of the ion dynamics, which makes them equivalent in the linear electrostatic collisionless regime, which is the one considered here. Additional features can be optionally switched on in some codes, like for example a non-circular geometry of the magnetic flux surfaces, or non-adiabatic models for the electrons.
The main basic difference of the three codes, even when circular flux surfaces are considered and the electrons are treated as adiabatic, resides in the numerical algorithm which is used to solve the model equations.
In fact, the Lagrangian algorithm is used for ORB5, the Eulerian algorithm for GENE, and the Semi-Lagrangian algorithm for GYSELA. This difference of the numerical schemes, makes the detailed cross-code comparison and verification against analytical theory even more meaningful - the numerical result is controlled not to depend on the numerical approximation of the basic model, but only on the considered physics.
The tests have been divided into two main classes, depending on the model used for the treatment of the electrons. In the first class, where the electrons are treated adiabatically, analytical dispersion relations exist in literature, and this makes not only a cross-code benchmark, but also a detailed verification of the codes, possible. On the other hand, when the electrons are treated kinetically, no analytical theory presently exists, and therefore a cross-code benchmark only has been performed.

The first test with adiabatic electrons has been chosen in a regime  where all three codes can be compared, namely with a magnetic equilibrium with circular flux surfaces. The frequency and damping rates of GAMs have been observed to fit well among codes, in the limit of moderate-low values of the safety factor, and for small values of the wave-number normalized to the ion Larmor radius (see Sec.~\ref{sec:adele_broad_q}).
In the same regime, a comparison with the analytical dispersion relation of Zonca-1996~\cite{Zonca96}, where no FOW effects are retained, and with the explicit formulas for the frequency and damping rate respectively of Sugama-2006~\cite{Sugama06} and Sugama-2008~\cite{Sugama08}, where FOW effects are retained to the first-order, has also been successfully done.
When introducing a non-circular geometry of the flux surfaces, the codes ORB5 and GENE have also been been benchmarked and verified against the analytical dispersion relation of Gao-2009~\cite{Gao09}, for a scan on the flux surface elongation. The result has been a good agreement of the codes for both frequency and damping rates, a quantitative agreement with the analytical theory for the frequency, and qualitative for the damping rates (see Sec.~\ref{sec:adele_broad_elo}).

When a regime with higher radial wave-numbers is considered, the ion FOW effects play a more important role. A comparison of ORB5, GENE and GYSELA with adiabatic electrons, with the analytical theories of Sugama-2008, Qiu-2009 and Gao-2010 has been made, scanning in the range $ 0 < k_r \rho_i \le 0.45$ (see Sec.~\ref{sec:adele_loc_modes}).
All codes have shown a very good comparison of the frequency with each other for all values of $k_r \rho_i$, and a good comparison with the analytical theory for low values of $k_r \rho_i$. The difference with the analytical theory which is found for higher values of $k_r \rho_i$, is thought to be due to the breaking of the regime of validity of the analytical theories, derived as expansions for small values of wave-numbers. The damping rate has given a very good matching of all codes, especially at moderate and large values of $k_r \rho_i$, where the theories of Qiu-2009 and Gao-2010 have been recovered.
At low values of $k_r \rho_i$, corresponding to low values of the damping rate, a difference among codes has been found, due to the general difficulty to measure low damping rates. For example, for a PIC code like ORB5, a high resolution in number of markers is necessary to kill the statistical noise and properly measure a very low value of damping rate, but typically some uncertainty still remains, unless a very big number of markers is used.

Benchmark tests with kinetic electrons have also been performed, with ORB5 and GENE, in a low-$k_r\rho_i$ regime, with circular flux surfaces, and moderate value of the safety factor. The results of the two codes have been found to fit very well. No analytical theory presently exists providing the modification of the frequency and  damping rate due to the effect of the kinetic electrons, therefore no verification has been possible in this regime. The scan of the frequency and damping rate in the ion to electron mass ratio has shown that a convergence of the frequencies with the analytical prediction obtained with adiabatic electrons is found when electrons are sufficiently light (when approaching realistic values of $m_i/m_e$ for hydrogen and deuterium plasmas) whereas no convergence is found for the damping rate, which stays one order of magnitude higher than the analytical prediction obtained with adiabatic electrons (in agreement with Ref.~\cite{Zhang10}).

Detailed convergence tests have been performed with all three codes in order to assess the numerical stability for the considered GAM dynamics. Convergence scans have been done for ORB5 with respect to the number of markers, which characterizes the type of discretization of a PIC code (see Appendix~\ref{sec:appendix_ORB5}). Analogously, the numerical description of the simulations performed with GENE and the convergence scans in $v_\|$ are reported in Appendix~\ref{sec:appendix_GENE}. Finally the description of the numerical parameters used for simulations with GYSELA, and convergence scans in the spatial and velocity space are presented in Appendix~\ref{sec:appendix_GYSELA}.

In conclusion, we have made a choice of three gyrokinetic codes and tested them for the physics of linear electrostatic collisionless GAMs in different regimes, by means of verification against analytical theory and cross-code benchmarks. These tests have shed light on the regimes of validity of the different analytical theories derived in the different limits. In particular, we have shown that there is not one approximate analytical formula, which can be applied for all the considered regimes. In fact, each considered formula has been found to match the results of the numerical simulations in a different regime of application, but to fail in other regimes. These regimes have been properly identified here, making their usage more sensible for the future. These tests have also improved the trustability of the codes. In particular, we have shown that the results of the three selected codes match very well for all simulations performed in regimes where the damping rate is moderate or high, whereas some 
differences have been found for very small values of the damping rates, where the numerical error can strongly affect the measurement. These tests performed on zonal structures like GAMs, and complementary tests performed on the linear dynamics of microturbulence modes (see for example Ref.~\cite{Goerler16}), serve to prepare a solid basis for a more comprehensive theoretical understading of the turbulent transport in tokamak plasmas, based on the numerical simulations with the set of available gyrokinetic codes, analytical theory, and intermediate reduced models, which is one of the major goals of our community.

\section*{Acknowledgements}
This work has been carried out within the framework of the EUROfusion Consortium and has received funding from the Euratom research and training program 2014{\textbackslash}-2018 under Grant Agreement No. 633053, for the WP15-ER-01/IPP01 project on ``Verification and development of new algorithms for gyrokinetic codes'', and WP15-ER-01/IPP02 project on ``Micro-turbulence properties in the core of tokamak plasmas: close comparison between experimental observations and theoretical predictions''. The views and opinions expressed herein do not necessarily reflect those of the European Commission.
This work was also partly supported by a grant from the Swiss National Supercomputing Centre (CSCS) under project IDs s617 and s704.  Simulations were performed on the International Fusion Energy Research Center (IFERC) CSC Helios supercomputer within the framework of the VERIGYRO project, and on the CINECA Marconi supercomputer within the framework of the OrbZONE project. The authors acknowledge discussions with F. Zonca, P. Lauber, E. Poli, D. Del Sarto, A. Ghizzo, P. Niskala, \"O. G\"urcan, P. Morel. Part of this work was done while two of the authors, A. Biancalani and I. Novikau, were visiting LPP-Palaiseau (France), whose team is acknowledged for the hospitality.

\begin{appendices}

\section{Numerical convergence tests with ORB5}
\label{sec:appendix_ORB5}

The numerical stability of the codes is crucial to be investigated, in order to assess the efficiency and the regime of validity of the algorithm. A test consists in measuring the convergence of GAM frequencies $\omega_{GAM}$ and damping rates $\gamma_{GAM}$ with different number of markers. This kind of tests has been done for simulations with adiabatic electrons and repeated for simulations with kinetic electrons (see Sec.~\ref{sec:ORB5-model} for a description of these two models for the electrons).

For these scans a plasma configuration with a major radius $R_0 = 1.3$ m and a minor radius $a=0.13$ m was chosen with the toroidal magnetic field on axis be $B=1.9$ T and flat profiles for the safety factor, with $q=3.5$, temperature (defined from the value of $\rho^* = 1/160 = 0.00625$) and density profiles (with value irrelevant for the present electrostatic simulations). Electrostatic linear simulations are evaluated with an initial electric potential perturbation with $k_r=2\pi/a$. This configuration corresponds to the case depicted in Fig.~\ref{fig:omegagamma_q} (point with $q=3.5$), and Fig.~\ref{fig:FKE-omegagamma_me} (points with $m_i/m_e=2000$ and $m_i/m_e=4000$).

For the simulations with adiabatic electrons, the typical spatial grid is $(\rho,\alpha_1,\alpha_2) = 64\times64\times4$ and the time step is 40 $\Omega_i^{-1}$ (where the ion cyclotron frequency is evaluated here with the magnetic field on axis, i.e. at $\rho=0$, and with $\alpha_1$ and $\alpha_2$ being the two periodic coordinates, i.e. the poloidal and toroidal angles). Simulations with $1000$ time-steps are considered, with a total time length of $4 \cdot 10^4$ $\Omega_i^{-1}$, where we observe about 10 GAM oscillations. 
A scan in the number of ion markers is performed, from $10^5$ to $10^8$. For kinetic electrons, the typical spatial grid is $(\rho,\alpha_1,\alpha_2) = 64\times64\times4$, and the time-step is $2$ $\Omega_i^{-1}$. For the case with kinetic electrons, simulations with $2 \cdot 10^4$ time-steps are considered,  a number of ion markers of $10^8$, and the number of electron markers is scanned from $10^7$ to $5 \cdot 10^8$.


\begin{figure}[t!]
\begin{center}
\includegraphics[width=0.47\textwidth]{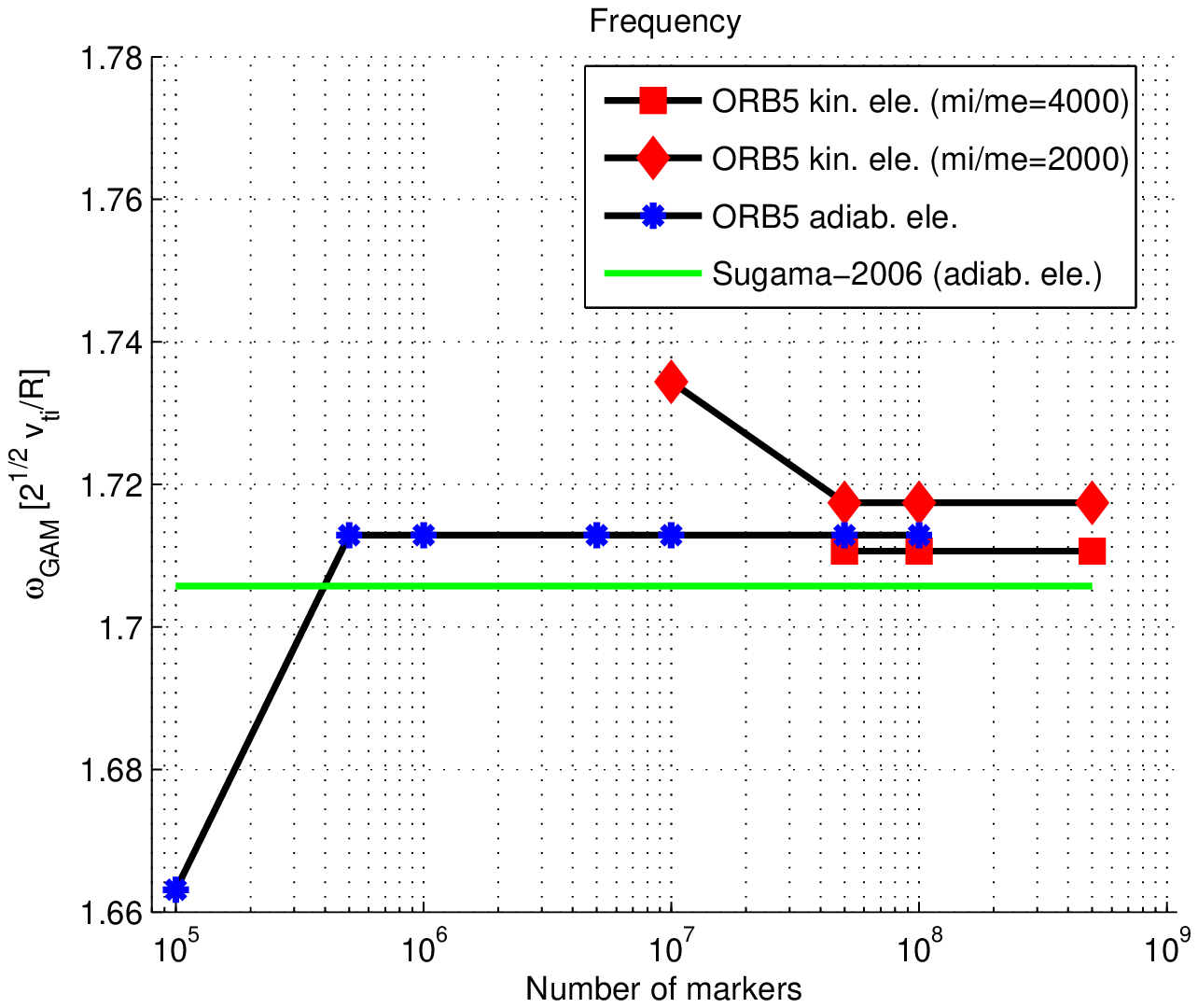}
\includegraphics[width=0.47\textwidth]{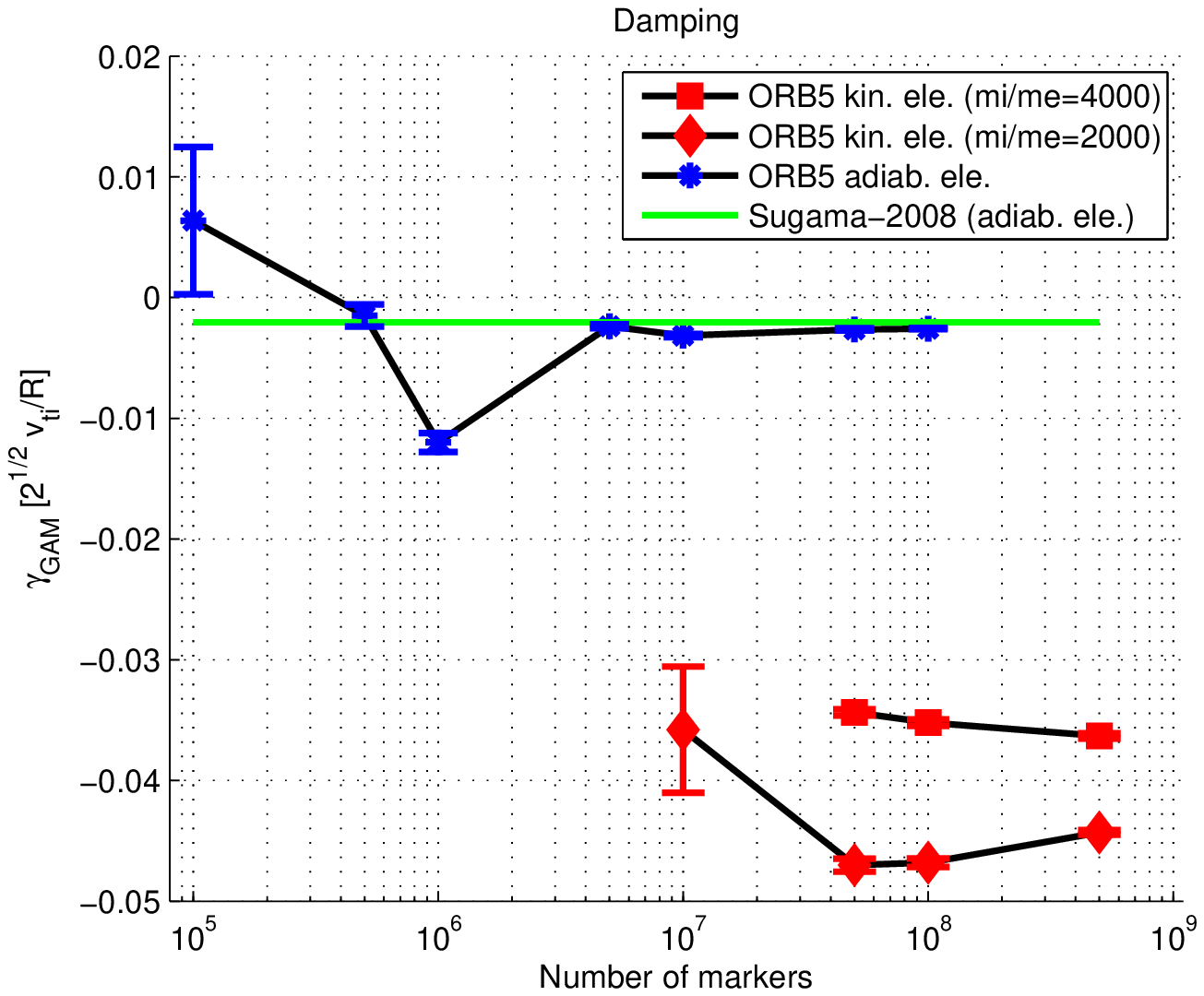}
\vskip -1em
\caption{\label{fig:appendix}GAMs frequencies (left) and damping rates (right). The number of markers for the ions (for simulations with adiabatic electrons, depicted by blue stars) and for the electrons (for simulations with kinetic electrons, depicted by red squares and diamonds) is indicated on the horizontal axis. Error bars are also indicated for the values of the damping rates.}
\end{center} 
\end{figure}

The frequency has been calculated directly by measuring the averaged period of oscillation at one radial position. To apply other techniques, like for example the Fourier decomposition, it's necessary to have more oscillations that increases significantly the calculation time of simulations. The damping rate and its standard deviation have been found by using the method of least squares.
Results of the convergence tests are given in Fig.~\ref{fig:appendix}, where it can be seen that for the case of adiabatic electrons, the frequency and the damping rate converge well to the analytical value calculated by using the explicit expressions of Sugama-2006 and Sugama-2008~\cite{Sugama06,Sugama08}, for increasing number of ion markers.
On the other hand, the absolute values of the damping rates for simulations with kinetic electrons are found to stay considerably higher (as described in Sec.~\ref{sec:kin_ele}), and no convergence with the results of simulations with adiabatic electrons is observed in the range of number of electron markers considered. The GAM frequency does not change much with the  number of markers, except for the cases with very small number of markers. Error bars are also reported in the values of the damping rates, to emphasize that at very low number of markers, the Monte-Carlo error becomes comparable with the physical signal damping.

In the simulations with kinetic electrons performed with ORB5 and discussed in this paper, the dynamics of passing electrons is treated kinetically, and consequently high frequency oscillations are observed on top of the lower frequency GAM oscillation (see also Ref.~\cite{Biancalani14}). These high-frequency oscillations correspond to the limit of kinetic Alfv\'en waves for $\beta$ going to zero (electrostatic model) at fixed temperature, also known as the $\omega_H$-mode~\cite{Lee87}. These high-frequency oscillations have been observed to create numerical instabilities for low number of markers (below $10^7$). For this reason, the results of simulations with kinetic electrons and electron markers below $10^7$ have not been reported in Fig.~\ref{fig:appendix}.

Regarding the numerical parameters of the simulations of GAMs with broad radial structure described in Sec.~\ref{sec:adele_broad_eq}, we have used a spatial grid of ($N_\rho$,$N_\theta$,$N_\phi$) = 256x64x4 and a time step of $dt=$100 $\Omega_i^{-1}$, with $N_i=10^8$ markers. The length of the simulations is $4\cdot 10^5 \, \Omega_i^{-1}$, corresponding to 400 time steps.
Regarding the simulations of GAMs with fine radial structure described in Sec.~\ref{sec:adele_loc_eq}, a typical simulation has a spatial grid of ($N_\rho$,$N_\theta$,$N_\phi$) = 256x64x4 and a time step of 25, 50 and 100 $\Omega_i^{-1}$, with $10^7$ and $10^8$ markers.


\section{Numerical convergence tests with GENE}
\label{sec:appendix_GENE}

GENE simulations are carried out considering an initial density
perturbation with the same sinusoidal functional form as described in Sec.~\ref{sec:adele_broad_eq}. In order to match the radial wave-number
of the initial perturbation, the radial domain $L_x$ is adapted for each value
of $k_r=2\pi/L_x$. The mid-radius  location, $r/a=0.5$ is the reference
position used to measure all normalization quantities and define the
dimensionless machine size parameter $\rho^*$. The typical resolution used in
the radial direction is one point per ion larmor radius, with the number of
points adapted such as to have always one grid-point located at  $r/a=0.5$. A
high spatial resolution is used in the parallel direction, up to 96 points, which turn
out to be necessary in order to correctly converge the GAM damping for the
large $q$ - small $k_r$ cases.  In velocity space we consider the domain
$L_{v_\|}\times L_{\mu}$=$6\times12$, a choice that will be justified in the
following. A typical grid is $n_{v_\|}\times n_{\mu}=256\times32$ points, where
the high resolution in the parallel velocity is motivated by the need of
avoiding the recurrence problem. A detailed discussion of this issue is
outside the scope of this paper, and the interested reader is referred to
e.g.~\cite{Pueschel-PhD}, where the recurrence problem is discussed
in details. With the aforementioned resolution, the recurrence time is longer
than the final simulated time in all cases considered here, thus the mode
frequency and damping can be easily extracted. Alternatively one could have
used a small hyperdiffusion in the $v_\|$ direction obtaining the same result. However, in general, we prefer avoiding introducing any numerical dissipation as this might impact the residual level of zonal flows (not considered in this paper).

\begin{figure}[t!]
\begin{center}
\includegraphics[width=0.45\textwidth]{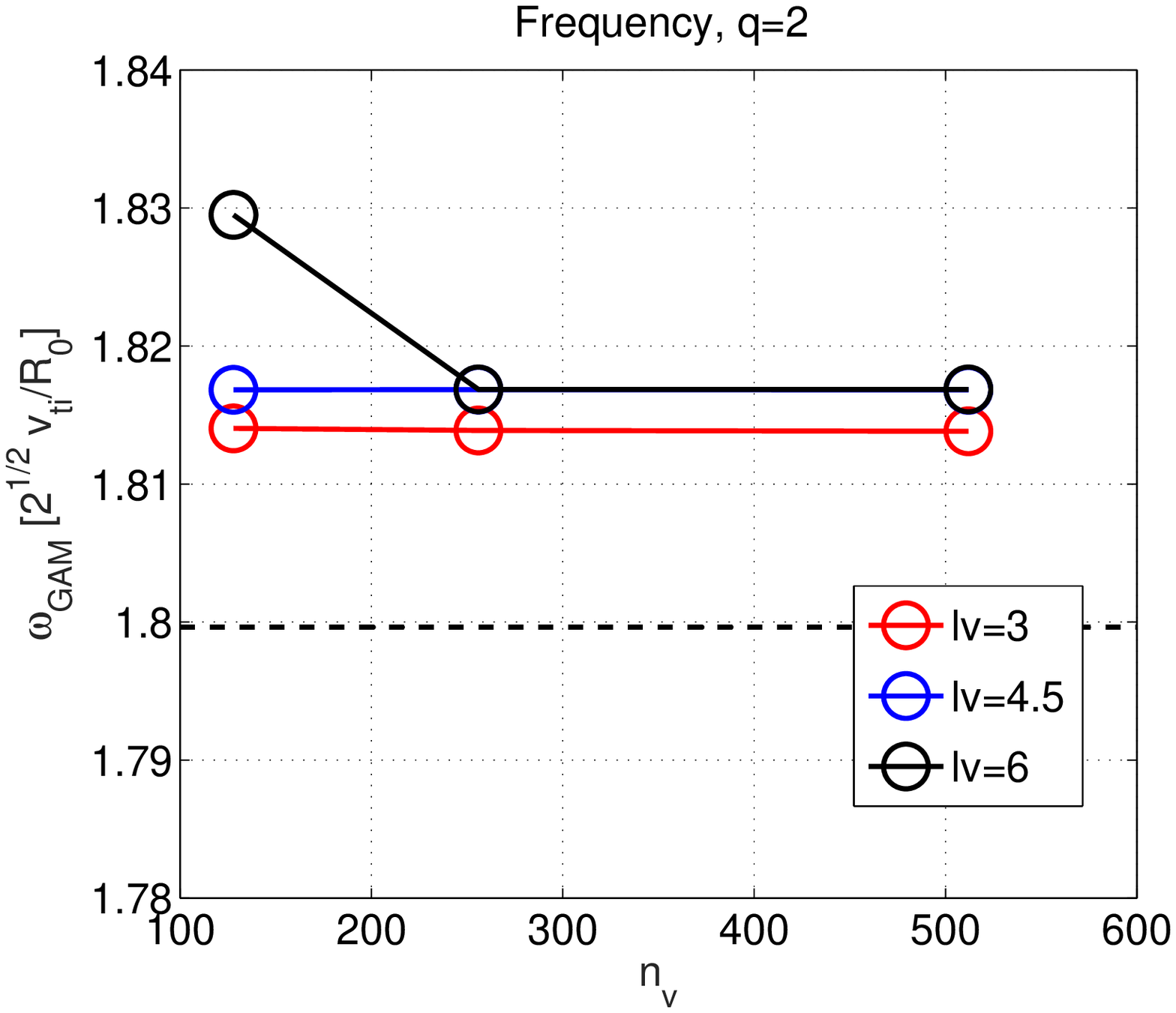}
\includegraphics[width=0.45\textwidth]{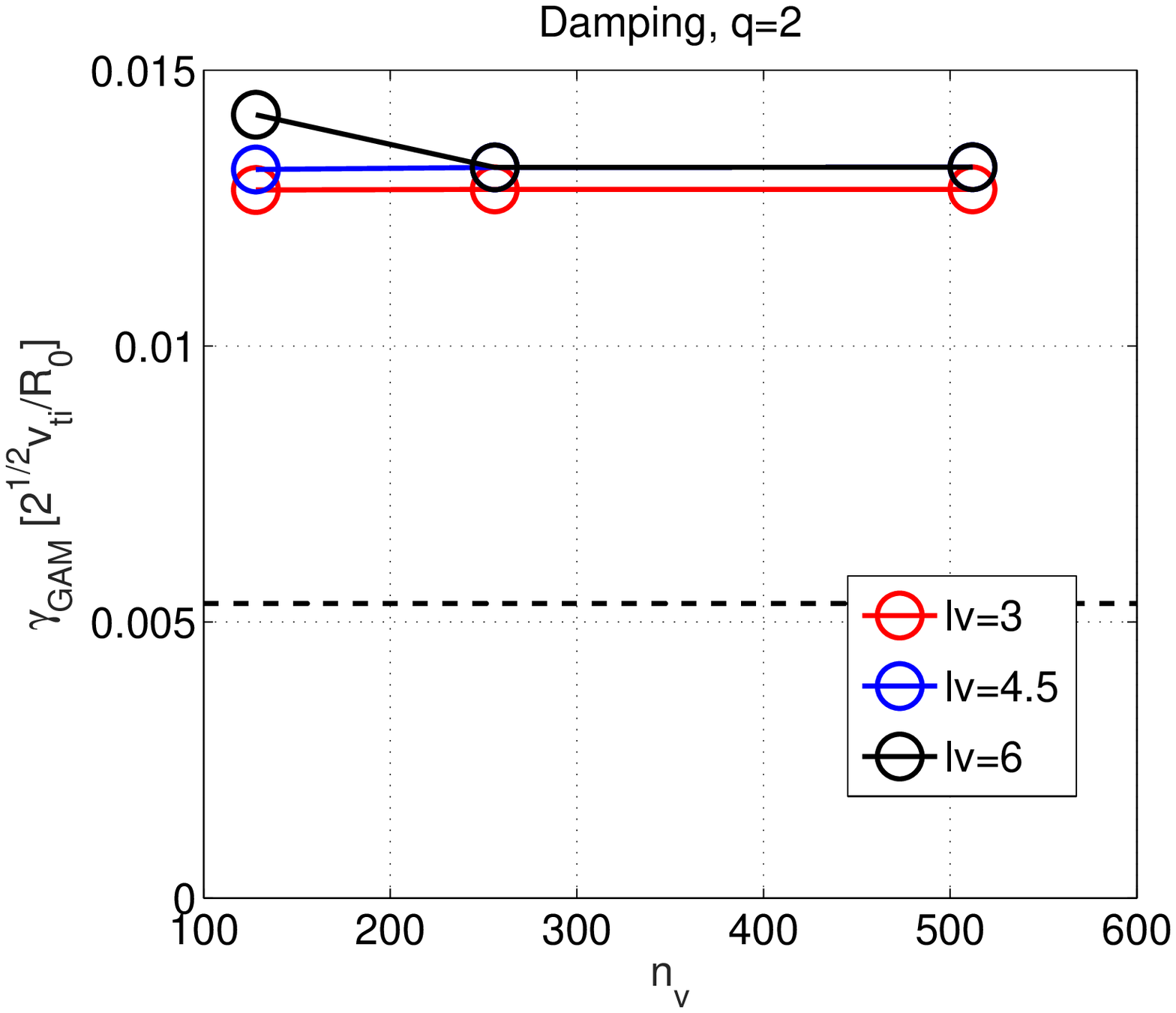}\\
\includegraphics[width=0.45\textwidth]{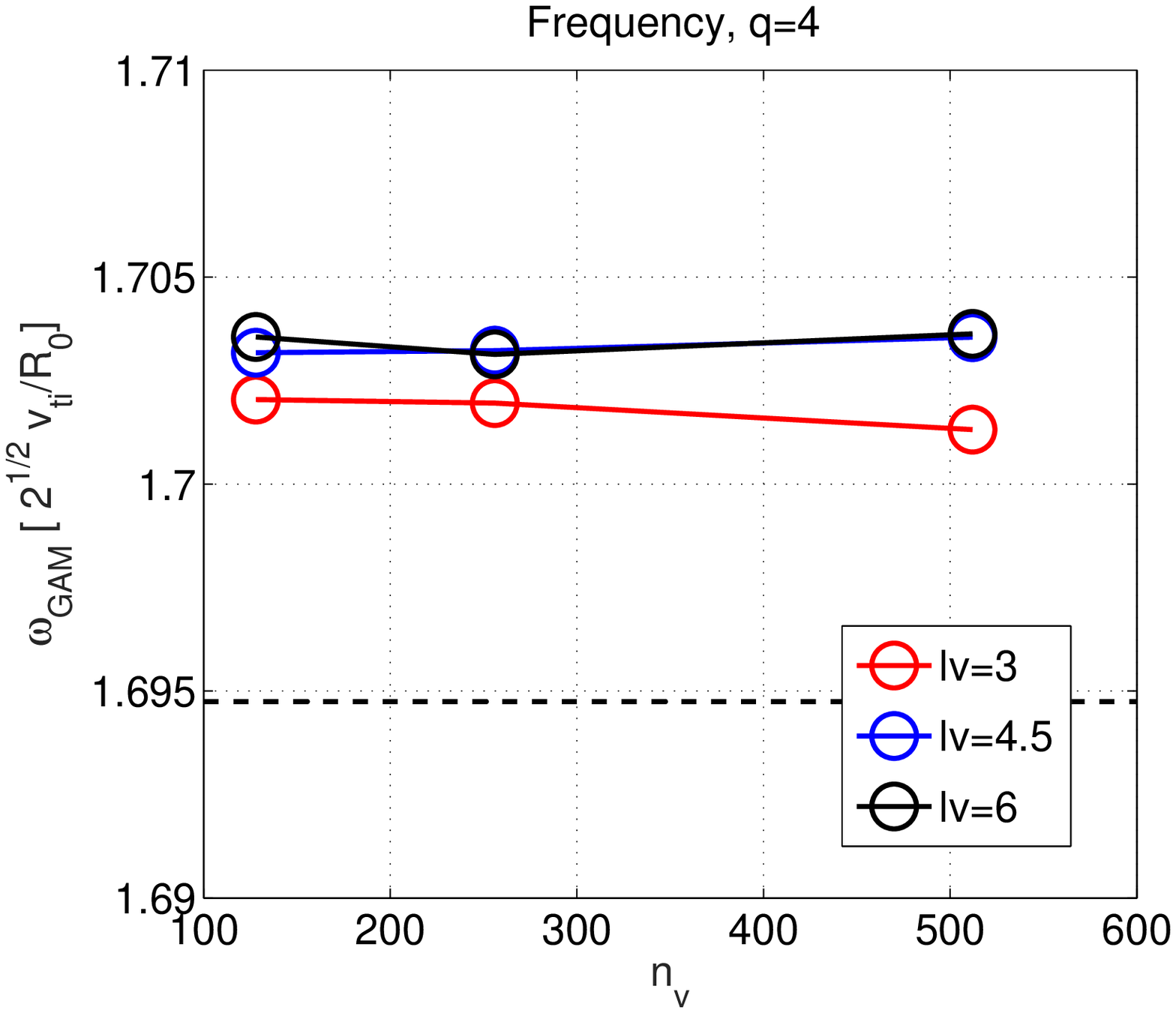}
\includegraphics[width=0.45\textwidth]{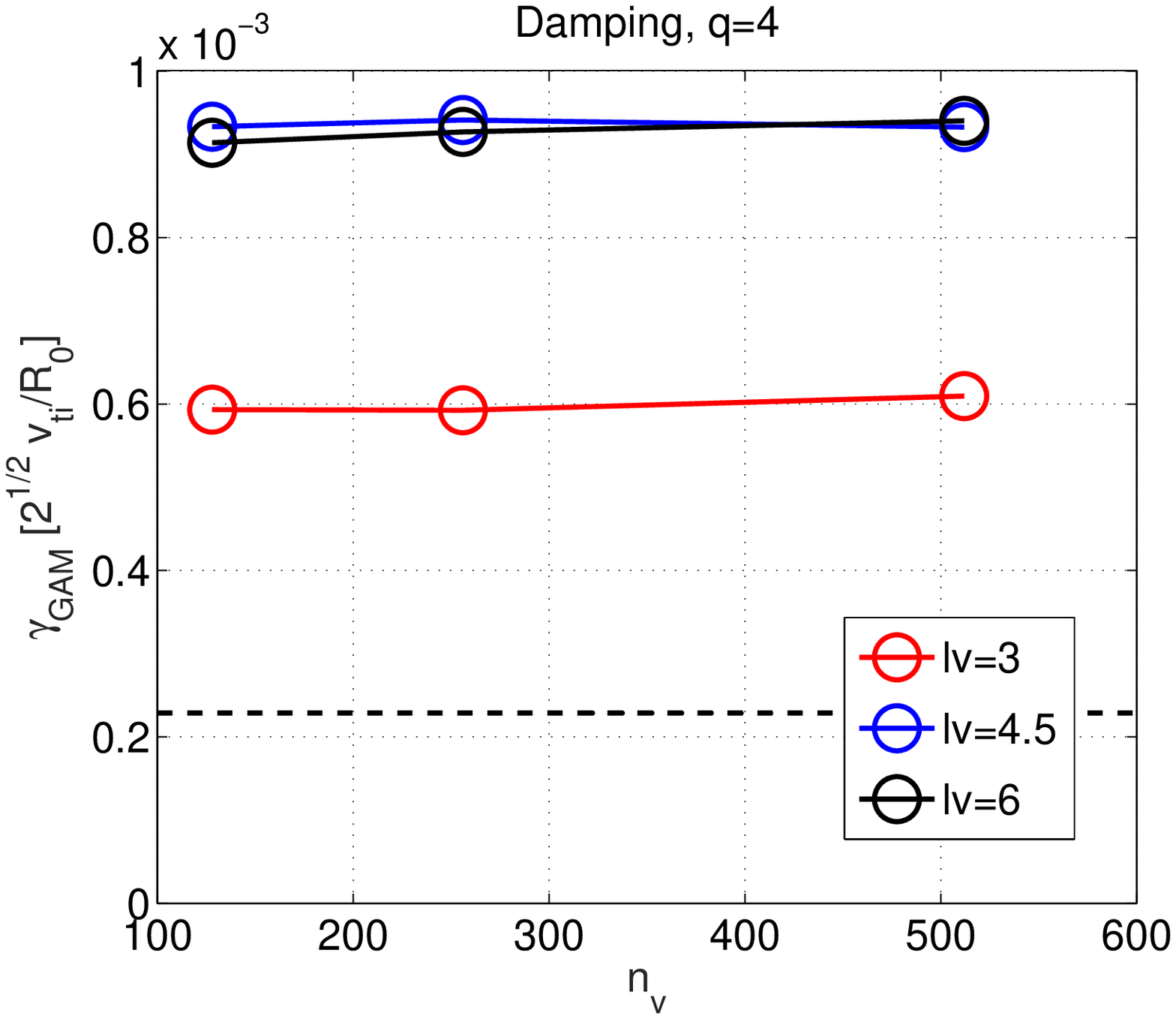}
\vskip -1em
\caption{
\label{fig:GENE_flat_q_conv}GAM frequency (left) and damping (right) for different resolutions and extension along $v_\|$ direction. At the top, the results for $q=2$, and at the bottom for $q=4$. The analytic prediction of Sugama-2006 and Sugama-2008 is reported with a dashed black line. Results obtained with flux-tube version of GENE with adiabatic electrons.
}
\end{center} 
\end{figure}

The properties of the GAM are evaluated by analyzing the time traces of the
flux-surface-averaged electrostatic potential $\bar\phi$,
measured at mid-radius. When comparing flux-tube and global simulations, the
same time interval is used. Simulations with adiabatic electrons are run
typically up to 150 $R_0/ \sqrt{2} \, v_{ti}$, in order to collect sufficiently long statistics (for the large damping cases it suffices to run the simulation for
much shorter times). The damping rate $\gamma_{\rm GAM}$ of the GAM is then
evaluated by separately fitting maxima and minima of the curve
$\bar\phi$ in time.  The frequency is computed using an Hilbert transform. 
 
In Figure \ref{fig:GENE_flat_q_conv} we plot GAM frequency and damping for two different values of the safety factor (2 and 4 respectively) varying the ion velocity space domain and
resolution. 
These simulations have been performed with the flux-tube version of GENE with adiabatic electrons. They have been repeated for global simulations (results not shown here) obtaining, as expected, the same behavior and an almost perfect agreement with local results.
We observe how the GAM frequency rapidly converges, whereas the damping is much more sensitive to resolution, and a
sufficiently large velocity space must be considered in order to correctly
converge the simulation results. 

We remark that kinetic electron runs are instead carried out for a significantly
shorter time, $\sim 30 R_0/ \sqrt{2} \, v_{ti}$, as the damping is found to be much stronger and it is therefore not necessary to simulate longer times.

\section{Numerical convergence tests with GYSELA}
\label{sec:appendix_GYSELA}

\begin{figure}[b!]
    \centering
\includegraphics[width=7.cm]{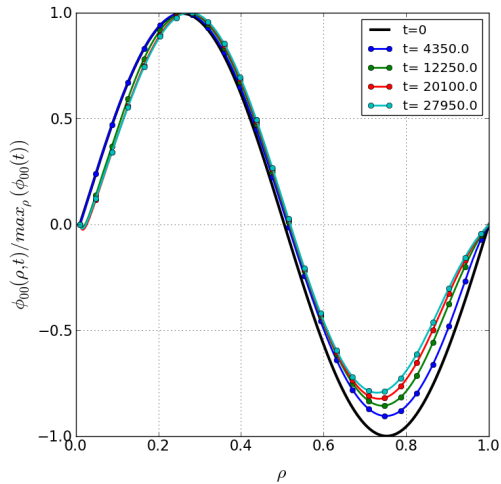}
    \caption{Time evolution of the radial profile of the zonal component $\bar\phi$ for the initial time (black line) and $4$ different times ($t=4350\,\Omega_i^{-1}$, $t=12250\,\Omega_i^{-1}$, $t=20100\,\Omega_i^{-1}$ and $t=27950\,\Omega_i^{-1}$. $\rho$ is the normalized radial position, i.e $\rho=(r-\rmin)/L_r$.}
    \label{fig:krrhoi_Q2}
\end{figure}

The convergence scan proposed for GYSELA has been performed with the same parameters as described in Sec.~\ref{sec:adele_broad_eq}. Density and temperature profiles are considered flat. The flat safety factor is taken equal to $2$ for the following tests. Electrons are considered adiabatic. 
In GYSELA, due to its full-$f$ character, the initial condition is performed on the distribution function $F_s$ and consists of an equilibrium distribution function $F_{s0}$ added to a perturbation $\delta F_{s}$, namely $F_s = F_{s0} + \delta F_s$. 
Then, the electrostatic potential $\phi(r,\theta,\varphi)$ is computed at time $t=0$ by solving quasi-neutrality equation \eqref{eq:GYSELA_QN}. 
In the present test, the perturbation part $\delta F_s$ reads $\delta F_s = F_{s0}\: g(r)$ with 
$g(r)=-\frac{1}{r}\left(k_r\cos(k_r r)-k_r^2 r\sin(k_r r)\right)$ 
where $k_r=(k+1)\pi/L_r$ with $k\in \mathbb{N} $, and $L_r=160\rho_s$. 
The corresponding radial profile of the zonal component $\phi_{00}(r)\sim \sin(k_r r)$ is plotted in Fig.~\ref{fig:krrhoi_Q2} (black line) for $k=1$ which is the value used for the following simulations.

In GYSELA, the $5$D space $(r,\theta,\varphi,\vpar,\mu)$ is uniformly discretized with $\Nr\times\Ntheta\times\Nphi$ points in the 3D real space and $\Nvpar\times\Nmu$ points in the $2$D velocity space. 
This mesh grid is fixed in time with $r\in[0,L_r]$, $\theta\in[0,2\pi]$, $\varphi\in[0,2\pi]$, $\vpar\in[-\alpha\, v_{T_i},\alpha\, v_{T_i}]$ and $\mu\in[0,L_\mu]$.
Due to the toroidal axisymmetry of the test the number of toroidal points $\Nphi$ is fixed to $\Nphi=8$. 
A comparison (not presented here) with $\Nphi=16$ has shown really good agreement with $\Nphi=8$. Simulations with $\Nphi=4$ would be probably close to those with $\Nphi=8$ but are not possible in the code due to parallelization constraint. This technical constraint could be removed. However simulations with so little number of points in toroidal direction are not standard simulations, so choice has been made to run with $\Nphi=8$ and to postpone the required modification of the code for now.
The maximum of thermal velocities in parallel velocity space is fixed at $\alpha=7$.
A simulation with $\alpha=5$ has been performed (not presented here) showing very small departure ($<2\%$) compared to the case $\alpha=7$. 
However, as this value could have more impact for larger $q$ values due to resonance position the value $\alpha=7$ has been preferred for the following tests.   
$L_\mu$ is fixed to $L_\mu=12\, T_i/B_0$ (with $B_0=1$). 
All simulations have been performed for a flat safety factor profile equal to $2$ and until $t=50000\,\Omega_{i}^{-1}$. Flat density and temperature profiles are also considered with $\tau_e=T_e/T_1=1$.

\begin{figure}[t!]
    \centering
     \includegraphics[width=7.cm]{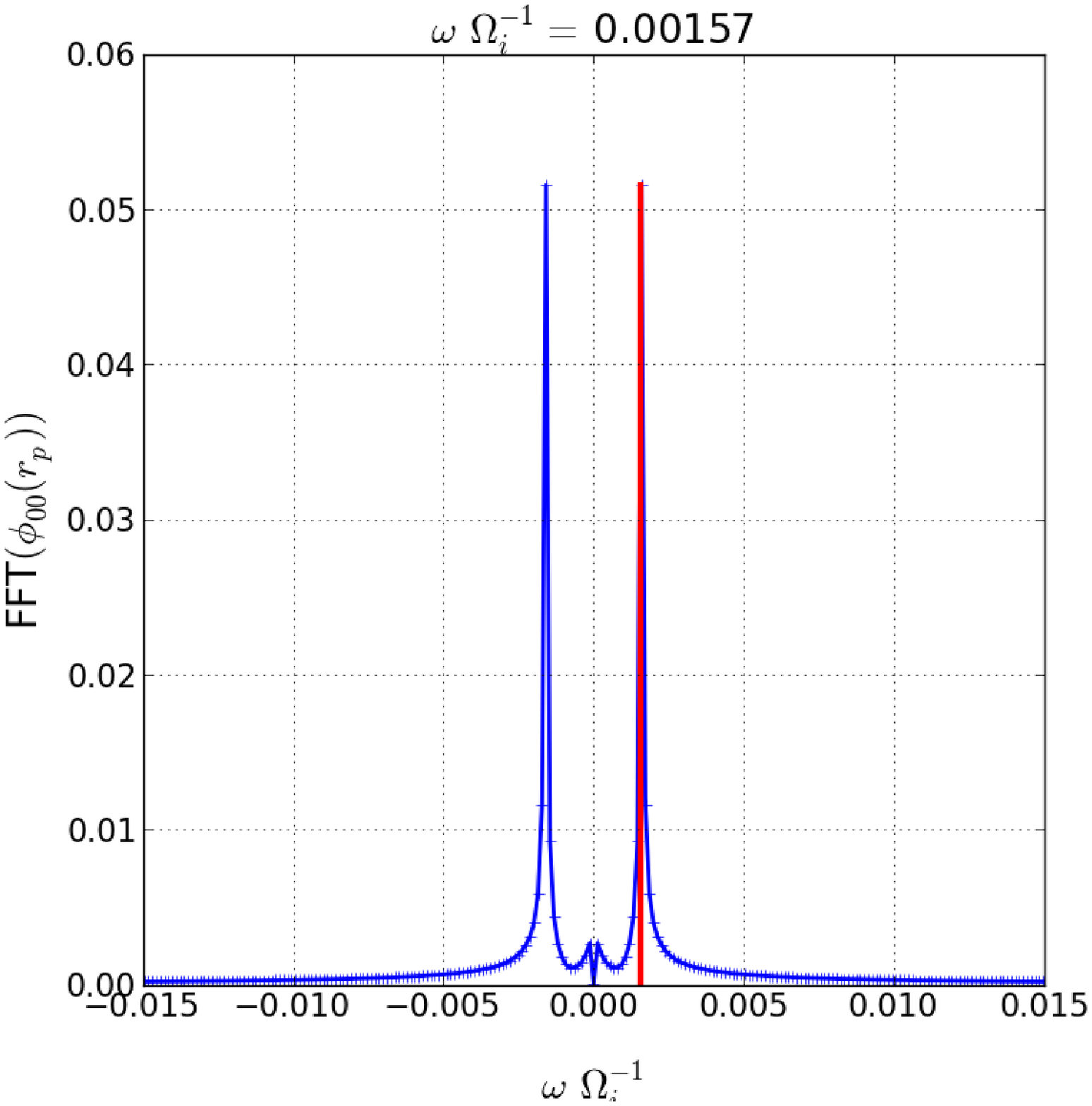}
    \includegraphics[width=7.cm]{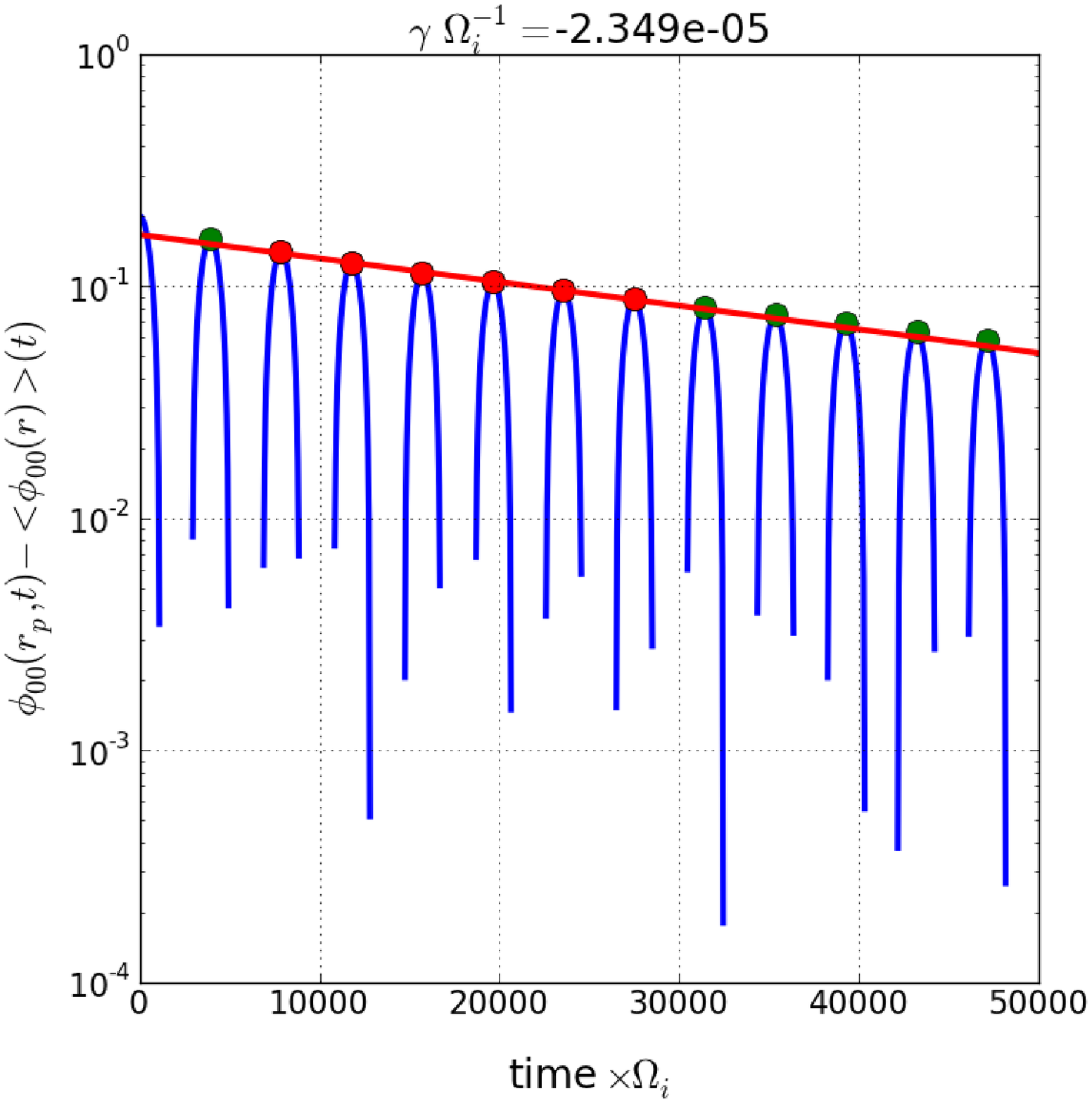}
    \caption{On the left, Fourier transform in time of $\bar\phi(r=r_p)$ used to compute the frequency with $r_p$ the radial position of the maximum value of $\bar\phi(r)$ at initial time. On the right, time evolution of $\bar\phi(r_p,t)-<\phi_{00}(r)>_r(t)$ used to compute the damping rate. $r_p$ is the radial position of the maximum value of $\bar\phi(r)$ at initial time. The green points correspond to the maximum values. The $6$ red points correspond to the points used for the linear interpolation (red line).}
    \label{fig:omegagamma_Q2}
\end{figure}

\begin{table}[b!]
  \centering
  \begin{tabular}[h]{c|c|c|c|c|c|c|c|c}
    case   & $N_r$ & $N_\theta$ & $N_{\vpar}$ & $N_\mu$ & $\Delta t\,\Omega_i$ & $k_r\rho_i$  & $\gamma\,\Omega_i^{-1}$     & $\omega\,\Omega_i^{-1}$\\
    \hline
    1      & $256$ & $64$       & $128$       & $32$    & $5.$                 & $0.05630867$ & $0.02664329$                & $1.81126121$ \\
    2      & $256$ & $64$       & $128$       & $32$    & $10.$                & $0.05630867$ & $0.02663145$                & $1.81126121$ \\
    3      & $256$ & $64$       & $128$       & $32$    & $25.$                & $0.05630868$ & $0.02657794$                & $1.81126121$ \\
    4      & $256$ & $64$       & $128$       & $32$    & $50.$                & $0.05630874$ & $0.02653612$                & $1.81196116$ \\
    5      & $256$ & $64$       & $128$       & $16$    & $25.$                & $0.05630058$ & $0.02657711$                & $1.81126121$ \\
    6      & $256$ & $64$       & $128$       & $8$     & $25.$                & $0.05601206$ & $0.0265876$                 & $1.81126121$ \\
    7      & $128$ & $64$       & $128$       & $32$    & $25.$                & $0.05649958$ & $0.0266591$                 & $1.81126121$ \\
    8      & $512$ & $64$       & $128$       & $32$    & $25.$                & $0.0562041$  & $0.02655708$                & $1.81126121$ \\
    9      & $256$ & $256$      & $128$       & $32$    & $25.$                & $0.05630868$ & $0.02657654$                & $1.81126121$ \\
    10     & $256$ & $64$       & $64$        & $32$    & $25.$                & $0.05630868$ & $0.02655779$                & $1.81196116$ \\
    11     & $128$ & $64$       & $64$        & $8$     & $50.$                & $0.05620426$ & $0.02661145$                & $1.81126121$ \\
  \end{tabular}
  \caption{$11$ simulations performed for $q=2$ with $\Nphi=8$ by varying the number of points in $r$, $\theta$, $\vpar$ and $\mu$ directions. Results are compared on the radial wave number $k_r\rho_i$, the damping rate $\gamma$ and the frequency $\omega$ of the zonal component of the electrostatic potential.}
  \label{tab:test_conv}
\end{table}

Parameters and results are summarized in Table \ref{tab:test_conv}. 
Comparisons are performed on the three quantities: (i) the radial wave number $k_r\rho_i$, (ii) the damping rate $\gamma$ and (iii) frequency $\omega$ of the zonal component of the electrostatic potential $\phi_{00}$.
The radial wave number is computed with the following formula:
\begin{equation}
  \label{eq:krrhoi}
  k_r\rho_i=\rho_i\sqrt{\sum_{i=0}^{N_r}\left(\bar\phi_{\rm norm}(r_i,t)-\langle\bar\phi_{\rm norm}\rangle_r\right)^2}\bigg/\sqrt{\sum_{i=0}^{N_r}\left(\frac{d}{dr}\bar\phi_{\rm norm}(r_i,t)\right)^2}
\end{equation}
with $\bar\phi_{\rm norm}(r_i,t)=\bar\phi(r_i,t)/\max_{r_i}\bar\phi(r_i,t)$.
The values reported in Table \ref{tab:test_conv} correspond to the mean values of $k_r\rho_i$ computed at times where $\log(\bar\phi(r_p,t))$ is maximum with $r_p$ the radial position of the maximum value of $\bar\phi(r)$ at initial time.
The damping rate is estimated by using the method of least squares also on the maximum values of $\log(\bar\phi(r_p,t))$.  $\gamma$ values reported in Table \ref{tab:test_conv} are computed with 6 maximums (see red circles in Figure \ref{fig:omegagamma_Q2}).
Four first simulations (cases 1 to 4 in Table \ref{tab:test_conv}) have been performed for the same 5D mesh of $\sim 536.8$ millions of points $(\Nr,\Ntheta,\Nphi,\Nvpar,\Nmu) = (256,64,8,128,32)$ but varying the time step $\Delta t$ from $\Delta t=5\,\Omega_i^{-1}$ to $\Delta t=50\,\Omega_i^{-1}$. 
All the other simulations except the last one (cases 5 to 10) have been performed with $\Delta t\Omega_i=25$ varying: (i) the number of points in $\mu$ direction (case 5: $\Nmu=16$, case 6:$\Nmu=8$); (ii) the number of points in radial direction (case 7: $\Nr=128$, case 8: $\Nr=512$); (iii) the number of points in poloidal direction (case 9: $\Nr\times\Ntheta=256^2$) and (iv) finally the number of points in parallel velocity space (case 10: $\Nvpar=64$). 
The last case (case 11) corresponds to a simulation where all varying parameters have been taken to their smaller tested value, namely $\Delta t=50\Omega_i^{-1}$, $\Nr=128$, $\Ntheta=64$, $\Nvpar=64$ and $\Nmu=8$.

Considering case 1 as the reference case, the maximum relative error is less than $1\%$ for $k_r\rho_i$ and $\omega$ estimations and less than $2\%$ for $\gamma$ (see Table \ref{tab:test_conv}).
As conclusion all these simulations, even the coarse grained one (case 11), are fully accurate.
However considering that these tests have been performed for a small safety factor value $q=2$ and a small radial wave number $k_r\rho_i\sim 0.056$ we could suggest to avoid parameters where we observe a small departure from the reference case, namely $\Delta t=50\Omega_i^{-1}$ and $\Nmu=8$. 
Then, more secure parameters for larger $q$ values or larger $k_r\rho_i$ values could correspond to those of case 5, namely a mesh $(\Nr,\Ntheta,\Nphi,\Nvpar,\Nmu)=(256,64,8,128,16)$ of $268.4$ millions of points with a time step of $\Delta t= 25\,\Omega_i^{-1}$. 
Such a simulation requires $2$ hours on $256$ cores for $2000$ time iterations compared to the coarse grained simulation which takes around $1$ hour on $64$ cores ($1000$ iterations).


\end{appendices}

\end{document}